%% file: main.tex
\documentclass[10pt,twocolumn,letterpaper]{article}

\usepackage{cvpr}              

\input{preamble}

\definecolor{cvprblue}{rgb}{0.21,0.49,0.74}
\usepackage[pagebackref,breaklinks,colorlinks,allcolors=cvprblue]{hyperref}

\def\confName{CVPR}
\def\confYear{2027}

\newcommand{\mname}{{M\"obiuString}}

\title{\mname: Structured String Art via Optimizing Generatrices\\on the M\"obius Strip}

\author{
Xinding Zhu\(^{1}\) \quad Hanni Bao\(^{1}\) \quad Xinye Yang\(^{1}\) \quad Qinyi Wang\(^{1}\) \quad Yuzhong Xu\(^{1}\) \quad Jiazhou Chen\(^{1\dagger}\) \\
\(^1\)Zhejiang University of Technology \\
{\tt\small cjz@zjut.edu.cn}
}

\makeatletter
\g@addto@macro\@maketitle{\input{figures/fig_teaser_0010011}}
\makeatother

\begin{document}
\maketitle
\input{sec/00_abstract}
\input{sec/01_introduction}
\input{sec/02_related_work}
\input{sec/03_geometry}
\input{sec/04_method}
\input{sec/05_experiments}
\input{sec/06_discussion}
{
    \small
    \bibliographystyle{ieeenat_fullname}
    \bibliography{references}
}

{
  \clearpage
  \appendix
  \input{sec/A_proofs}
  \input{sec/B_budget}
  \input{sec/C_implementation}
  \input{sec/D_loss_ablations}
}
\end{document}

%% file: preamble.tex
\usepackage[T1]{fontenc}
\usepackage{iftex}
\ifPDFTeX
  \usepackage[utf8]{inputenc}
\fi
\usepackage{mathtools}
\usepackage{bm}
\usepackage{microtype}
\usepackage{multirow}
\definecolor{cvprblue}{rgb}{0.21,0.49,0.74}
\usepackage[pagebackref,breaklinks,colorlinks,allcolors=cvprblue]{hyperref}

\DeclareMathOperator{\softplus}{softplus}

\newcommand{\ind}{\mathbf 1}
\newcommand{\normal}{\mathbf n}
\newcommand{\tangent}{\mathbf t}

%% file: figures/fig_teaser_0010011.tex
\begin{minipage}{\textwidth}
  \centering
  \captionsetup{type=figure}
  \includegraphics[width=\textwidth]{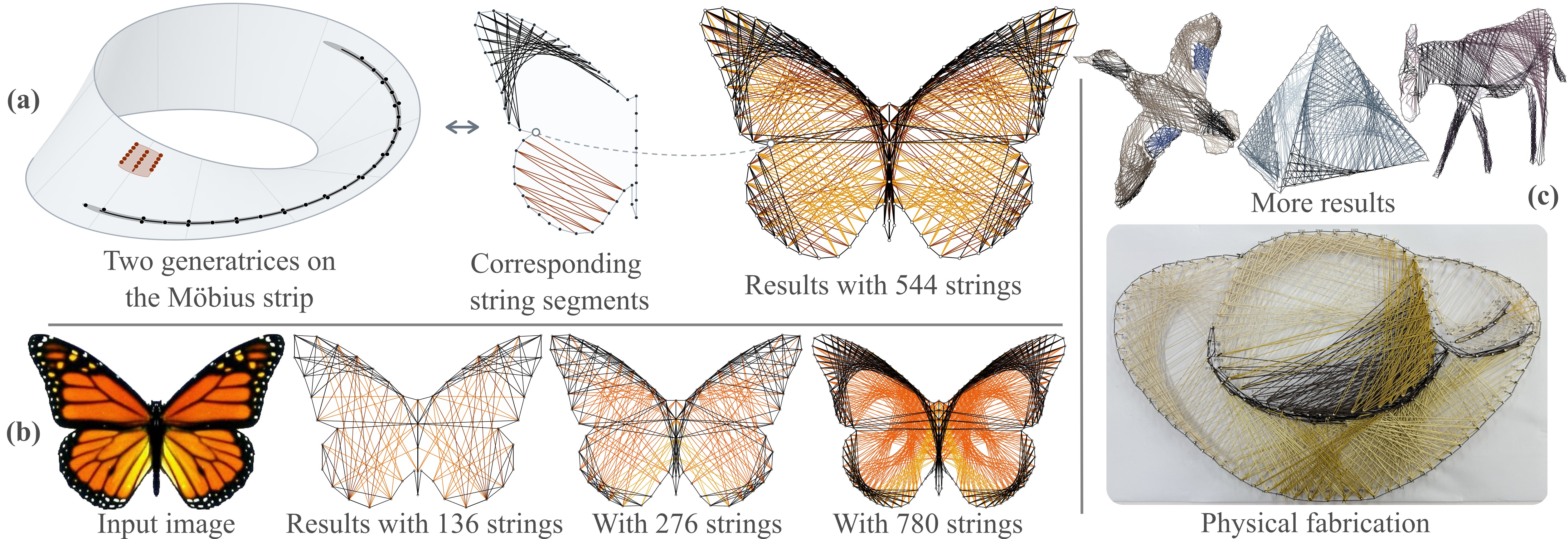}
  \caption{\textbf{Structured string art generated by M\"obiuString.}
(a) Generatrices on the M\"obius strip parameterize structured families of string segments, as illustrated by the wing inset. (b) Given an image, M\"obiuString can automatically generate string arts with different string budgets. (c) shows three more results (top) and a photograph of a handcrafted artwork with 987 strings in 4 colors (bottom).}
  \label{fig:teaser-0010011}
\end{minipage}
\par\vspace{\intextsep}

%% file: sec/00_abstract.tex
\begin{abstract}
String art is a worldwide handicraft that mimics an image with straight strings stretched between boundary pins. Existing image-driven methods primarily optimize individual strings for reconstruction fidelity, leaving the global organization of the strings weakly controlled.
In this paper, we introduce the generatrix on the M\"obius strip of unordered boundary-point pairs to formalize families of strings with finite generatrices, enabling explicit control over their arrangement.
Based on this new formalization, we propose M\"obiuString, a framework for generating structured, multi-color string art under a prescribed string budget. As an initialization, we place pins along boundaries of image-derived regions and select candidate string segments near each generatrix in the homeomorphic pin-index plane. During the iterative optimization, a 2DGS-based differentiable renderer jointly optimizes generatrix parameters and colors on multiple layers, while periodic discrete re-search revises string families to escape poor local configurations.
To our knowledge, M\"obiuString is the first automatic image-based method to generate string art with a structured string organization with freeform region boundaries. 
Evaluations with two datasets and a user study demonstrate the outperformance of M\"obiuString over baselines. 
\end{abstract}

%% file: sec/01_introduction.tex

\section{Introduction}
\label{sec:intro}
String art depicts an image with straight strings stretched between boundary pins, which is a popular handicraft all over the world. Strings are typically arranged in structured families with only small translational or rotational variations, creating a strong sense of spatial regularity. This regularity is a defining aesthetic characteristic of string art and is closely tied to its distinctive stretching fabrication process. Under a limited string budget, string segments with several different colors must contribute to the shape and appearance of the target image with layer occlusion. 
Therefore, generating string art is a widely-interested but challenging task that requires a structured organization of strings with multiple colors, while guaranteeing reconstruction fidelity.

Existing methods commonly construct string art incrementally by adding the string segment with the largest immediate reconstruction benefit~\cite{birsak2018,demoussel2022,je2019}. Even when candidates are evaluated against an image-wide objective, additive construction rarely revises earlier string segments; under tight budgets, early choices can consume strings needed elsewhere. Chord-based optimization~\cite{fang2016} evaluates multiple chords under a common objective, but does not provide a joint multilayer formulation for multi-color string art. Color extensions either optimize palette components independently and combine their arrangements~\cite{danalaStringArt,hachnochiStringArt} or incrementally select connection-color pairs~\cite{hachnochiStringArt}. These strategies do not jointly revise existing arrangements across color layers. Moreover, image fitting alone does not require strings to form structured groups, although parallel arrangements, gradual direction changes, and envelope-like patterns remain perceptually prominent in existing artworks.

In this paper, we present \textit{M\"obiuString}, an image-driven framework for generating structured, multi-color string art. Given a target image, a foreground mask, and a string budget, it derives regions and places pins along object and internal boundaries with budget-adapted spacing. Strings are organized into ordered monochromatic layers. Using differentiable vector rendering~\cite{li2020diffvg,liu2025bezier} based on 2D Gaussian Splatting, we render all layers as a complete composition and jointly optimize their connections and learnable colors under a shared image objective and a string-budget objective. 

To make string organization explicit, we introduce \emph{generatrices} in chord space. A chord is an unordered pair of boundary points; for a closed boundary, the chord space is topologically a M\"obius strip~\cite{morton1967}. A generatrix parameterizes a family of strings conveying a high spatial regularity. Our implementation uses finite generatrix segments in periodic pin-index coordinates and a seam-aware selection field to associate candidate chords with each family, supporting image-derived region boundaries of freeform shape, rather than the prescribed circular or rectangular frames assumed by prior methods, while preserving endpoint exchange and periodic continuity. Iterative refinement jointly updates generatrix parameters and string colors. Periodic discrete re-search (relocation, deletion, and regeneration) revises string families using binary-rendered evaluations, linking relaxed optimization to avoid local optima.

We evaluate M\"obiuString on two image datasets (Butterfly and FSS-1000) across different string budgets, comparing reconstruction quality, perceived organization, computational cost, and fabrication feasibility. Ablation studies show the validity of each component, and a user study with 24 participants demonstrates the superiority of M\"obiuString over existing SOTA methods. Handcrafted artworks with selected results further show the feasibility of physical fabrication.

Our contributions are:
\begin{itemize}
  \item \textbf{A generation framework M\"obiuString} that first achieves structured, multi-color string art with freeform, image-derived region boundaries. It jointly optimizes locations and colors of string segments through differentiable vector rendering based on 2DGS.
  \item \textbf{A new generatrix representation on the M\"obius strip} of non-oriented boundary-point pairs that explicitly formalizes the global structured regularity of string families.
\end{itemize}

%% file: sec/02_related_work.tex

\section{Related Work}
\label{sec:related}

String art generation integrates techniques spanning image approximation, differentiable computation, and geometric modeling. 
This chapter therefore reviews existing literature from three perspectives: string art generation methods, differentiable vector rendering algorithms, and mathematical foundations of the chord space.

\subsection{String art Generation}

String art approximates target images by selecting connections between boundary pins and overlaying string segments. Existing solutions mainly fall into two categories: greedy‑iterative selection and global least‑squares optimization with error‑diffusion sampling.

Greedy‑iterative solvers repeatedly select string segments that maximize residual gain and apply post‑processing to produce winding sequences for physical fabrication, which constitutes a widely adopted baseline paradigm for string art~\cite{birsak2018,demoussel2022,je2019,oleti2026,schouten2022}. Classic implementations under this paradigm combine greedy string selection with Eulerization to output fabrication‑ready connection graphs~\cite{birsak2018}.

The least‑squares branch first computes a fitness score for all candidate chords jointly and then samples chords via error‑diffusion sampling~\cite{fang2016}. While it maps chords onto a triangular parameter domain, it does not identify domain boundaries to form a M\"obius strip manifold and lacks a joint multi‑layer formulation for sparse color string art.

Other works focus on interactive geometric design, multicolor output demonstrations, and fabrication‑constraint modeling for string drawing~\cite{chuluunbat2026,chen2026spool,lo2019}.

Nevertheless, most of these methods treat individual string segments as basic units. They tend to produce dense string art patterns; when targeting well‑organized string segments, they can only impose indirect constraints through regularizers and lack the ability to directly control the global geometric structure of string segments at the representation level. With a small budget of strings, such approaches struggle to faithfully approximate the target image, and the resulting layout often appears unstructured and visually messy.

\subsection{Differentiable Vector Rendering}

Differentiable vector rendering acts as the key underlying component. It maps geometric primitives at a high level to the output of image space and builds differentiable loss mappings to enable gradient‑driven optimization for vector‑generation tasks.

DiffVG~\cite{li2020diffvg} pioneered general differentiable rasterization for vector graphics, enabling gradient-based optimization of vector shapes. Subsequent work extended Gaussian splatting to vector rendering: surface-oriented 2D Gaussian Splatting~\cite{huang2024} adapted Gaussian primitives for geometric surface representation, while NURBS Splatting~\cite{qiu2026nurbs} extended the splatting framework to NURBS curves. Bézier Splatting~\cite{liu2025bezier} achieves efficient differentiable rasterization by sampling 2D Gaussians along curves, which can realistically simulate the superposition of semi‑transparent strokes. This splatting‑based rendering paradigm directly inspires our imaging simulation of strings.

Furthermore, differentiable vector tools support diverse creation tasks including image vectorization~\cite{ma2022live}, diffusion‑driven vector sketch generation~\cite{arar2025swiftsketch,jain2023vectorfusion,vinker2022clipasso,xing2023diffsketcher}, and video vector drawing~\cite{zhu2026hypersketch,zhu2026dmt}. These studies demonstrate the feasibility of placing structured parameters before the differentiable renderer. Motivated by this insight, our pipeline takes high-level structured parameters as input to the rendering module to realize differentiable optimization for pin‑constrained straight chords.

\subsection{Mathematical Foundations of Chord Families}

Rich geometric theories underpin the parameterization of chord families. String segments are discrete chord collections, while chord families denote sets of chords in continuous chord space. Mathematical definitions and derivations are typically formulated for chord families. Morton~\cite{morton1967} proved that the second symmetric product of circle boundary unordered point pairs is topologically homeomorphic to a M\"obius strip, and this topological result forms the core starting point for our chord‑space modeling.

Extensive research has investigated the geometric relationships between chords and their envelopes, including analyses of string art envelope properties~\cite{quenell2009}, support‑function‑based line‑family constructions~\cite{dhar2025,sir2008}, as well as derivations for toroidal oriented chord families and diverse analytic chord family instances~\cite{bouthillier2018,fujimura2017,halbeisen2021,herr2026,kocik2013}.

The Hough transform establishes a duality mapping between image‑space lines and points in parameter space~\cite{duda1972,illingworth1988,hough1962}, such that sets of lines correspond to continuous curves within the parameter domain. These theorems and derivations concerning chords inspire our treatment of string segments in string art. We adopt the M\"obius topology as the generation domain for chord families, leverage Hough duality to characterize chord families' geometric behaviors, and further introduce engineering‑friendly endpoint‑index charts to construct a complete modeling system that balances theoretical analysis and practical computation.

%% file: sec/03_geometry.tex

\section{Parameterization in Chord Space}
\label{sec:geometry}
Instead of selecting individual string segments one by one as in prior methods, we organize them as structured families tailored to the artistic characteristics of regularity. We further find that a structured family of string segments can be Hough-transformed to a generatrix in the chord space, topologically equivalent to a M\"obius strip, as shown in Fig.~\ref{fig:chord-space-correspondence}(a). 

In this chapter, we first define this chord space mathematically in Sec.~\ref {sec:mobius}, then introduce the pin-index coordinate system for practical optimization in Sec.~\ref{sec:indexchart}.

\input{figures/fig_chord_space}

\subsection{Chord Space as a M\"obius Strip}
\label{sec:mobius}
Given a target object in image space, we denote its closed boundary as $S^1$, a sequence of pins $\{p_i\}$. A chord is an unordered pair of pins, such as $(p_i,p_j)$ and $(p_j,p_i)$, which refer to the same chord. Collecting all unordered pairs of points along the closed boundary $S^1$ yields a chord space:
\begin{equation}
 \mathcal M=\operatorname{Sym}^2(S^1)
 =(S^1\times S^1)/\bigl((p_i,p_j)\sim(p_j,p_i)\bigr),
 \label{eq:symmetric}
\end{equation}
where \(\operatorname{Sym}^2(S^1)\) is the second symmetric product of \(S^1\). 

The chord space $\mathcal M$ is topologically equivalent to a M\"obius strip, forming its manifold~\cite{morton1967}. Topological equivalence means they admit mutual continuous deformation without tearing or gluing, a property describing the continuous space of all chords on the circle. Unlike an ordinary planar domain where generatrices would be cut off at the boundary, the M\"obius strip maintains continuous generatrices across its seam. For string art, pins are finite discrete samples, so available string segments correspond only to discrete lattice points on the M\"obius strip rather than the full strip. String segments with coincident endpoints are degenerate and excluded from rendering and optimization.

Within the chord space, a one-dimensional curve is called a generatrix, as the orange curves or blue lines in Fig.~\ref{fig:chord-space-correspondence}(a). In the ideal continuous setting, each point on a generatrix corresponds to exactly one string segment in the image space, so a generatrix directly defines a family of string segments. Using generatrices, we explicitly encode structured chords with low-dimensional parameters, which distinguishes our method from per-chord representation. We introduce the Hough‑coordinate system as a theoretical analytic tool to understand the envelope patterns induced by generatrices. Full definitions, envelope derivations, and examples of analytic generatrices are provided in Appendix~A.

\subsection{Engineering Pin-Index Coordinates}
\label{sec:indexchart}

Hough coordinates are suitable for analyzing the envelope geometry of chord families and defining continuous metrics, yet they are inconvenient for direct parameter optimization. To address this,  we introduce the pin‑index coordinate system(Fig.~\ref{fig:chord-space-correspondence}(b)), another coordinate chart for the same M\"obius chord space. This coordinate system maps generatrices from the chord space(Fig.~\ref{fig:chord-space-correspondence}(a)) to structured families of string segments in image space(Fig.~\ref{fig:chord-space-correspondence}(c)), and is tailored for parameterization and solving with discrete pins. We assign cyclic continuous indices to pins on the circular boundary, and a string segment is represented by an index pair $(i,j)$, which naturally encodes the equivalence rules of unordered endpoints and cyclic periodic translation. The topological space described by these index coordinates is identical to the symmetric product space in Eq.~\eqref{eq:symmetric}, while the distance metrics defined on them differ.

Fig.~\ref{fig:chord-space-correspondence}(b) visualizes the M\"obius twist via coordinate transformation: crossing the seam boundary flips the corresponding position, i.e., $(i,n) \mapsto (0,i)$ or $(0,j) \mapsto (j,n)$, where $n$ is the axis length for $i$ and $j$ in the index coordinate system. When a generatrix crosses the upper boundary in the triangular index diagram, it continues from the corresponding position on the left side to preserve its continuity on the M\"obius manifold. The diagonal $i=j$ corresponds to degenerate string segments with coincident endpoints.

We represent each generatrix as a finite segment:
\begin{equation}
S_m^{\mathrm{idx}}:\quad
j=k_m i+b_m,\qquad i\in[u_m,v_m]
\end{equation}
where $k_m\in\{-1,+1\}$ is the fixed slope, $b_m$ is the intercept, and $[u_m,v_m]$ specifies the finite extent. Fig.~\ref{fig:chord-space-correspondence}(b) visualizes the $\pm1$-slope generatrices in the index space, which are the forms adopted in our algorithm. We collect ten other types of generatrices from previous work in Appendix~A, which are not used in our method due to the trade-off between computational efficiency and robustness.

%% file: figures/fig_chord_space.tex
\begin{figure}
  \centering
  \includegraphics[width=\linewidth]{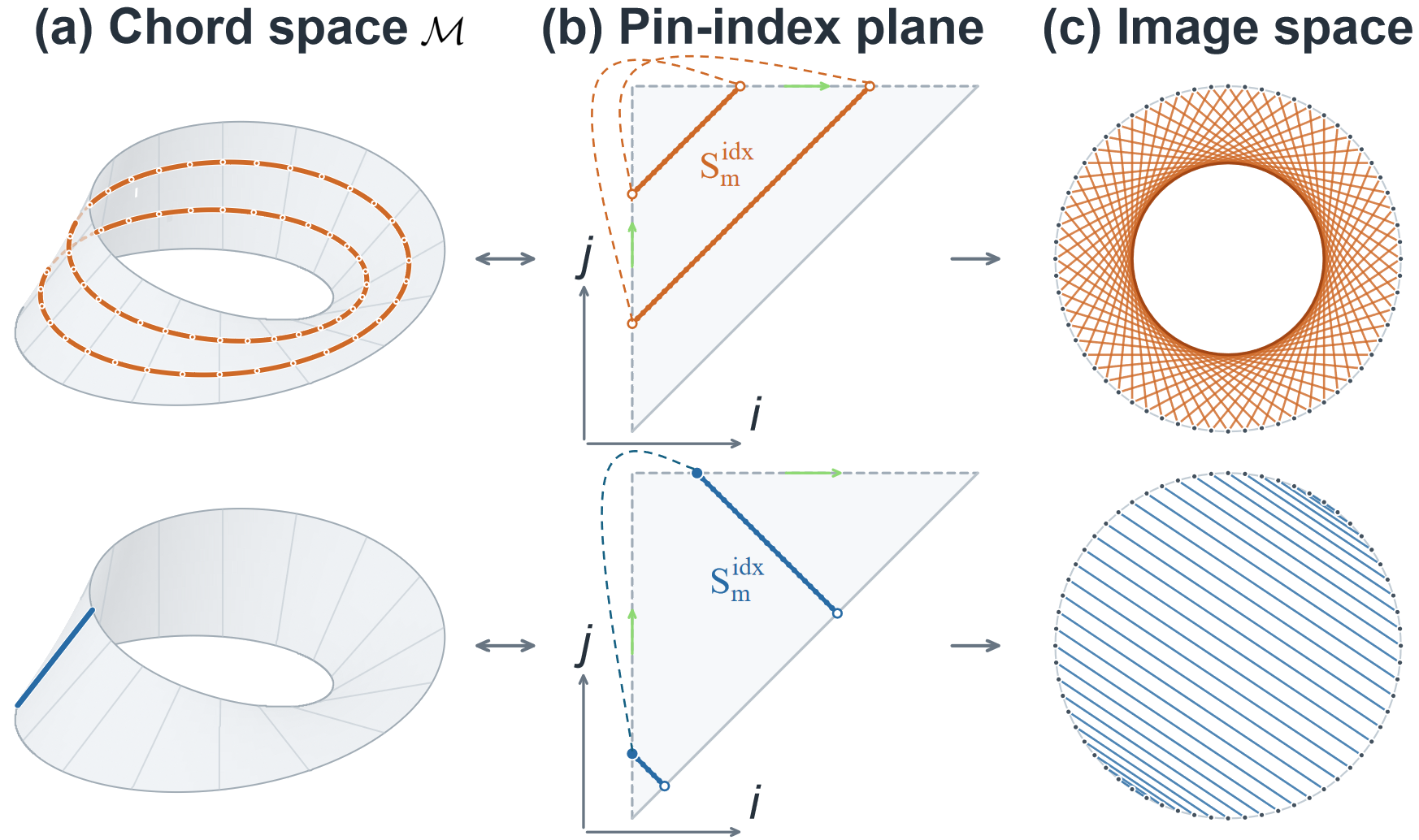}
 
  \caption{\textbf{The illustration of our parameterization in chord space.}
  (a) The continuous chord space $\mathcal{M}$ with a manifold of a M\"obius strip. (b) The discrete triangular pin-index plane with generatrices $S_m^\mathrm{idx}$; dashed curves illustrate the seam identification rule across the boundary of the triangular index plane, and green arrows show the direction on these boundaries. (c) The corresponding string patterns in the image space. The top and bottom rows show generatrices of slope $+1$ and $-1$, respectively. 
  } 
  \label{fig:chord-space-correspondence}
\end{figure}

%% file: sec/04_method.tex

\input{figures/fig_method_overview}

\section{Methodology}
\label{sec:method}
Given a target image $I^\star$, a foreground mask $M$, and a string budget $B$, we initialize regions and generatrices (Sec.~\ref{subsec:inti}), then alternate gradient refinement with discrete generatrix re-search (Fig.~\ref{fig:method-overview}). 

\subsection{Initialization}
\label{subsec:inti}

\noindent\textbf{Regions.}
We use a frozen SAM 2~\cite{ravi2024sam2} to generate region proposals. Region boundaries are evaluated with DINOv2~\cite{oquab2024dinov2} and Lab color features. The recursive subdivision maintains valid, hole-free regions and stops when further splitting brings no visual gain. Symmetric inputs are partitioned on one side and mirrored.

After region splitting, we assign color-layer slots to each region. A color layer is an independent drawing unit holding one string color and several generatrices, rendered inside-out to simulate occlusion from overlapping strings. We initialize a compact palette via Lab-space clustering~\cite{shi2024mondrian}: black outline is fixed, other string colors are learnable. Representing the target as color mixtures, we build material demand maps and select suitable color layers for each region group based on image-driven color demand.

\noindent\textbf{Pins.}
Boundaries of segmented regions are represented as piecewise cubic Bézier curves, where adjacent regions share the same curves. We sample pins along all these curves with approximately uniform arc‑length spacing following $h(B)=l_p\sqrt{l/B}$, where $B$ is the total string budget, $l$ denotes the diagonal length of the input image, and $l_p$ is a pin‑density parameter set to 26 in our experiments. The derivation of this spacing rule is provided in Appendix~B. Pins on shared boundaries are deduplicated while strictly preserving the cyclic ordering of boundary points.

\noindent\textbf{Generatrices.}
With region splitting and pin sampling done, we filter valid candidate chords: only pin-pair chords whose entire sampled segment lies within a single region are kept, while cross-region chords are discarded. Each valid chord is mapped to pin-index coordinates, and generatrix initialization is performed on this candidate set. Using each color layer's material demand map, we score every candidate chord and search for high-reward chord clusters in pin-index space, from which generatrix seeds are heuristically fitted, supporting straight generatrices with slopes $\pm 1$. Lightweight geometric filtering then discards generatrix slots that would yield many out-of-bounds or low-yield strings. This stage only produces heuristic initial guesses, not optimal solutions.

\subsection{Iterative Optimization}
\label{sec:diffrender}
This stage performs continuous-parameter iterative optimization with fixed generatrix slots. We update each generatrix's center and half-span through unconstrained endpoint parameters, keeping its slope fixed. Candidate chords receive Gaussian weights based on their point-to-segment distances to the generatrices, accounting for cyclic pin indexing and endpoint exchange. Within each color layer, overlapping families are combined by taking the maximum weight.

After obtaining each color layer's relaxed chord family, we adopt 2D Gaussian splatting for differentiable string rendering. Each string is modeled as a 2D Gaussian kernel with fixed width. Color layers are composited layer-by-layer in a predefined inside-out order, naturally handling occlusion and blending among strings and finally outputting the fully rendered string-art image.

We jointly optimize each generatrix's position on the pin-index plane and each color layer's thread color. Optimization is guided by the total loss $\mathcal L_{\mathrm{total}}$:

\begin{equation}
\mathcal L_{\mathrm{total}}
= \mathcal L_{\mathrm{vis}} + w_{\mathrm{cnt}} \mathcal L_{\mathrm{cnt}} + \mathcal L_{\mathrm{reg}}
\end{equation}

$\mathcal{L}_{\mathrm{vis}}$ is the visual loss for appearance and color.

\begin{equation}
\mathcal L_{\mathrm{vis}}
= w_{\mathrm{app}}\mathcal L_{\mathrm{app}}
+ w_{\mathrm{mis}}\mathcal L_{\mathrm{mis}}
+ w_{\mathrm{gain}}\mathcal L_{\mathrm{gain}}
+ w_{\mathrm{bg}}\mathcal L_{\mathrm{bg}}
\end{equation}

$\mathcal L_{\mathrm{app}}$ is the appearance loss, measuring the structural similarity between the rendered image $\boldsymbol I_{\text{render}}$ and the target $\boldsymbol I^\star$, including background regions.
\begin{equation}
\mathcal L_{\mathrm{app}} = 1 - \text{MS‑SSIM}(\boldsymbol I_{\text{render}},\boldsymbol I^\star)
\end{equation}




$\mathcal L_{\mathrm{mis}}$,  $\mathcal L_{\mathrm{gain}}$, and $\mathcal L_{\mathrm{bg}}$ target color‑related objectives. $\mathcal L_{\mathrm{mis}}$ is the mis‑color loss. It penalizes color discrepancies between rendered string layouts and the target image. $\mathcal L_{\mathrm{gain}}$ is the visible color‑gain loss. It encourages adequate string activation over foreground target regions. This loss applies a custom penalty to discourage insufficient visual contribution from individual strings. $\mathcal L_{\mathrm{bg}}$ is the background‑suppression loss. It prevents excessive strings from being placed over white regions requiring little to no string coverage in the input image, reserving empty space for the white background. Complete mathematical formulations for these three loss components are provided in Appendix~C.

$\mathcal L_{\mathrm{cnt}}$ is the chord-count budget loss, where $\widetilde K$ is the soft sum of chord soft gates $q_e$ (provided in Appendix~C) and $B$ is the target chord budget. It penalizes deviation from the desired chord quantity.
\begin{equation}
\mathcal L_{\mathrm{cnt}}
=
\left(\frac{\widetilde K-B}{B}\right)^2
\end{equation}

$\mathcal L_{\mathrm{reg}}$ is an auxiliary regularization loss for selection-gate binarization and color anchoring, where $C$ is the total number of color layers and $N$ denotes the number of candidate chords. It drives soft gates toward binary values and keeps thread colors anchored to initial palette values.
\begin{equation}
\mathcal L_{\mathrm{reg}}
= w_{\mathrm{bin}} \frac{1}{N}\sum_{e} q_e(1-q_e)
+ w_{\mathrm{col}} \frac{1}{C}\sum_{c=1}^{C}\big\|\boldsymbol c_c - \boldsymbol c_c^0\big\|_2^2
\end{equation}

\subsection{Generatrix Re‑search}
Subject to a pre‑fixed total number of generatrix slots, generatrix re‑search performs discrete structural adjustments to remedy continuous optimization’s limitation: it only fine‑tunes locally around active generatrices and may fail to escape poor initial layouts.

The pipeline adopts gradient‑based candidate scoring for low‑cost screening. A temporary continuous weight variable $\xi_e$ is introduced on top of the current binary chord set. The expected improvement $b_e$ of candidate chord $e$ is estimated via differentiation:
\begin{equation}
b_e=-\frac{\partial}{\partial \xi_e}
\left[\mathcal L_{\mathrm{vis}}(\xi)
+2\left(\frac{\sum_{e'}\xi_{e'}-B}{B}\right)^2\right]
\label{eq:probe}
\end{equation}
where $\mathcal L_{\mathrm{vis}}(\xi)$ denotes visible reconstruction loss under these temporary weights, and $\sum_{e'}\xi_{e'}$ is the corresponding soft string count. Larger values indicate larger expected loss reduction when activating the chord.

Guided by these gradient scores, we generate structural‑change proposals covering generatrix deletion, regeneration, relocation, and color swap. Each generated proposal first undergoes short‑range iterative optimization for fine‑tuning. After fine‑tuning, we evaluate each proposal using $\mathcal L_{\mathrm{total}}$. A proposal is accepted and generatrix states are updated if $\mathcal L_{\mathrm{total}}^{\mathrm{after}} < \mathcal L_{\mathrm{total}}^{\mathrm{before}}$.

\input{figures/fig_qualitative_comparison}

\subsection{Implementation Details}
We alternate Adam‑based gradient updates. We set learning rates to $0.005$ for generatrix‑family parameters, and $0.002$ for spool colors. The baseline executes $120$ total iterations with gradient norm clipped to $5$ and random seed $0$. Discrete search is periodically triggered between $20\%$–$80\%$ of training iterations to mitigate local‑optimum issues. In our total loss function, these weights are fixed as $w_{\mathrm{app}}=4$, $w_{\mathrm{mis}}=2.5$, $w_{\mathrm{gain}}=6$, $w_{\mathrm{bg}}=0.02$, $w_{\mathrm{cnt}}=2.0$, $w_{\mathrm{bin}}=0.02$, and $w_{\mathrm{col}}=0.03$ for all experiments. Further implementation details are provided in Appendix~C.


%% file: figures/fig_method_overview.tex
\begin{figure*}[t]
  \centering
  \includegraphics[width=\textwidth]{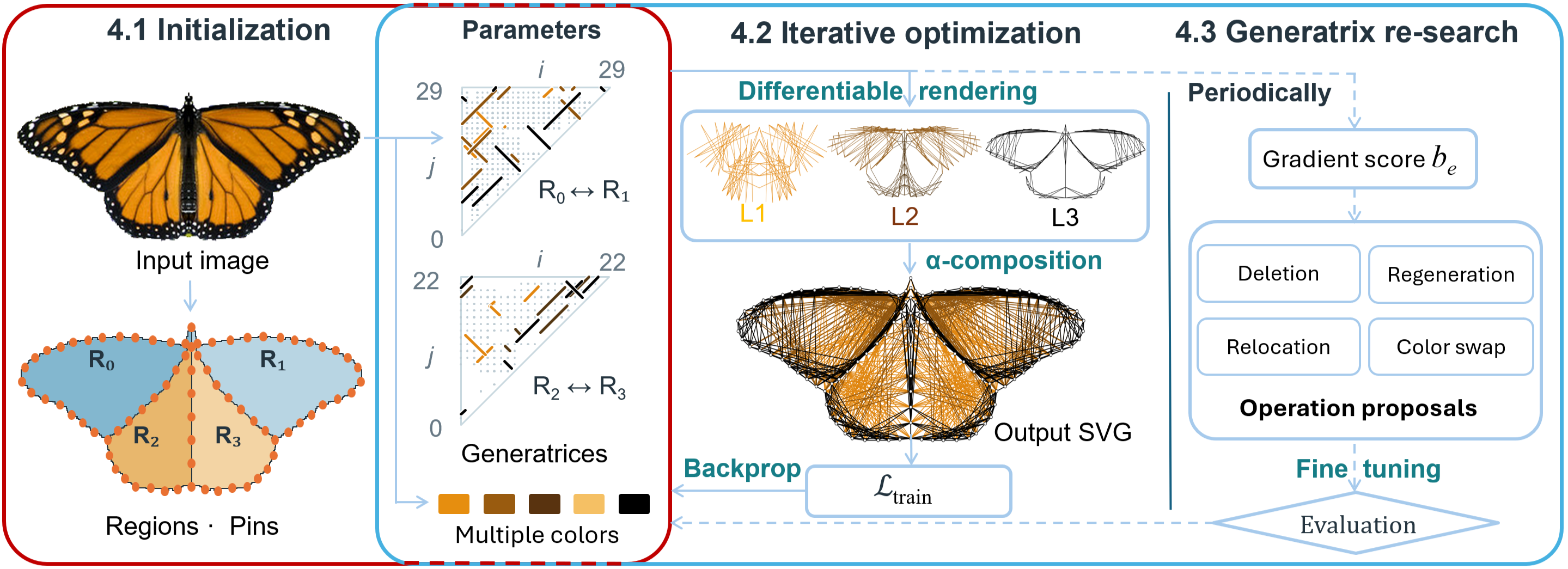}
  \caption{\textbf{Method overview}.
  Our framework starts from the initialized region segmentation and pin-indexed chord representation, and then iteratively updates generatrix parameters and colors simultaneously, guided by hierarchical differentiable rendering. Periodically, the generatrix re-search revises string families with 4 different operation proposals to further improve the robustness of our method.
}
  \label{fig:method-overview}
\end{figure*}

%% file: figures/fig_qualitative_comparison.tex
\begin{figure*}[t]
  \centering
  \begingroup
  \setlength{\tabcolsep}{0pt}
  \renewcommand{\arraystretch}{1}
  \def\qualPanelGap{8pt}
  \def\qualCellGap{2pt}
  \def\qualTargetWidth{38pt}
  \def\qualBudgetWidth{9pt}
  \edef\qualPanelWidth{\the\dimexpr(\textwidth-\qualPanelGap)/2\relax}
  \edef\qualCellWidth{\the\dimexpr(\qualPanelWidth-\qualTargetWidth-\qualBudgetWidth-\qualCellGap*5)/4\relax}
  \def\qualImageHeight{40pt}
  \def\qualRowGap{2pt}
  \newcommand{\qualResult}[3]{%
    \begin{minipage}[c]{\qualCellWidth}
      \offinterlineskip
      \hbox{%
        \vbox to \qualImageHeight{\vfil
          \hbox to \qualCellWidth{\hfil
            \includegraphics[width=\qualCellWidth,height=\qualImageHeight,keepaspectratio]{#1}%
            \hfil}\vfil}%
        \llap{\vbox to \qualImageHeight{%
          \hbox to \qualCellWidth{\hfil
            {\fontsize{7}{9}\selectfont $#3$}\hspace{1pt}}\vfil}}}%
      \vskip1pt
      \hbox to \qualCellWidth{\hfil
        {\fontsize{6}{7}\selectfont\strut
          \if\relax\detokenize{#2}\relax\else$K=#2$\fi}\hfil}
    \end{minipage}}
  \newcommand{\qualTarget}[2]{%
    \begin{minipage}[c]{\qualTargetWidth}
      \centering
      {\fontsize{7}{8}\selectfont\shortstack{Target from\\#2}}\par\vspace{3pt}
      \includegraphics[width=\qualTargetWidth,height=\qualTargetWidth,keepaspectratio]{#1}
    \end{minipage}}
  \newcommand{\qualBudget}[1]{%
    \makebox[\qualBudgetWidth][c]{%
      \rotatebox[origin=c]{90}{\fontsize{7}{9}\selectfont $B=#1$}}}
  \fontsize{8}{9}\selectfont
  \hbox to \textwidth{%
  \begin{minipage}[t]{\qualPanelWidth}
    \centering
    \begin{tabular}{@{}c@{\hspace{\qualCellGap}}c@{}}
      \qualTarget{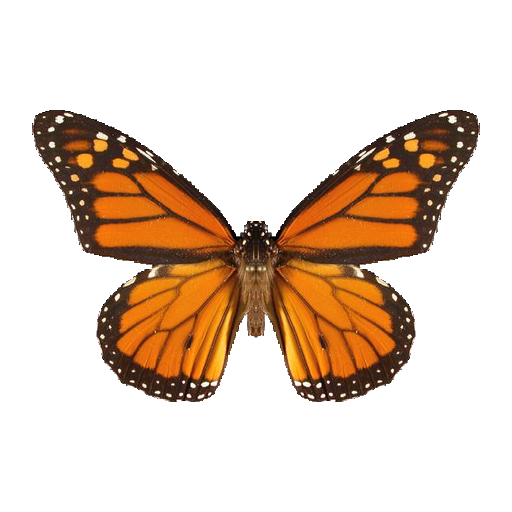}{Butterfly} &
      \begin{tabular}{@{}c@{\hspace{\qualCellGap}}*{3}{c@{\hspace{\qualCellGap}}}c@{}}
        & \textbf{Ours} & DanAla & SpiderPrinter & Roy (color) \\[2pt]
        \qualBudget{128} &
        \qualResult{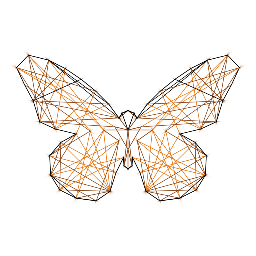}{126}{} &
        \qualResult{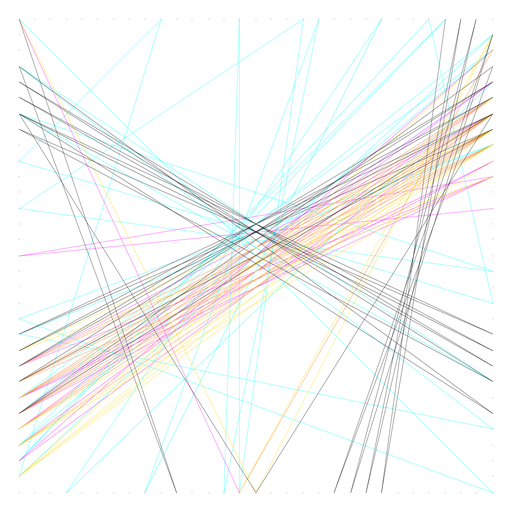}{}{} &
        \qualResult{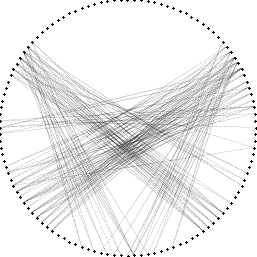}{}{} &
        \qualResult{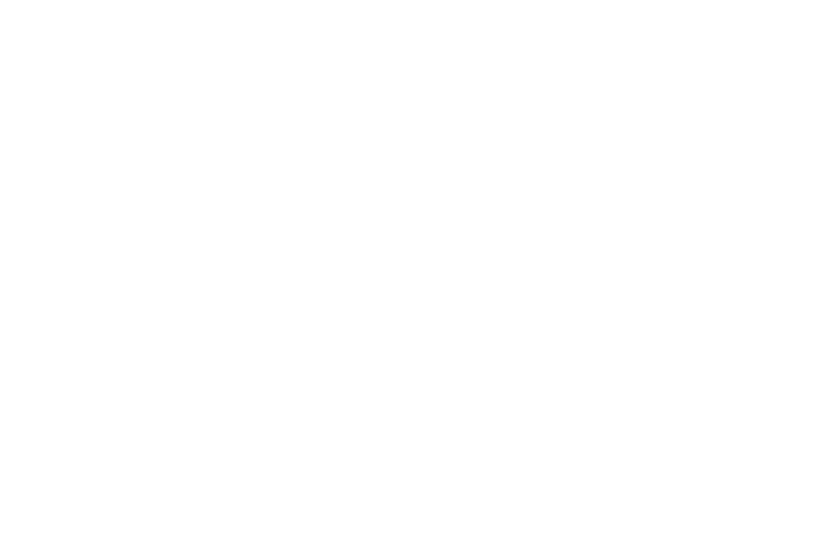}{}{\dagger} \\[\qualRowGap]
        \qualBudget{256} &
        \qualResult{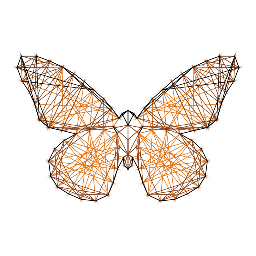}{270}{} &
        \qualResult{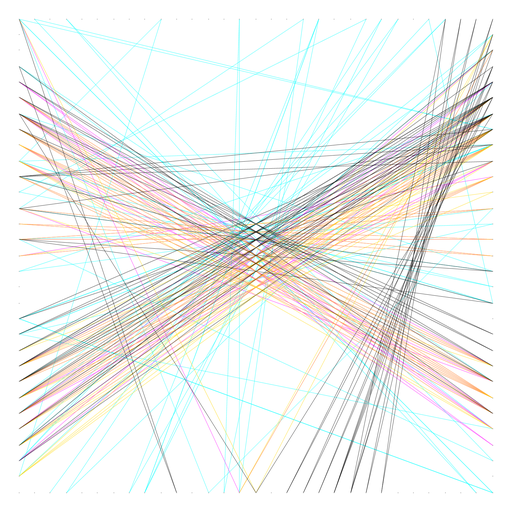}{}{} &
        \qualResult{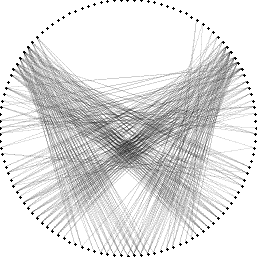}{}{} &
        \qualResult{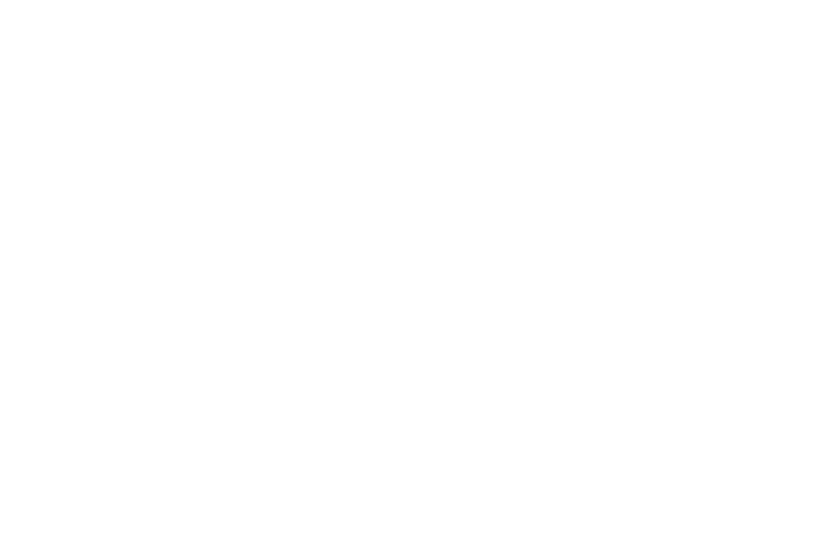}{}{\dagger} \\[\qualRowGap]
        \qualBudget{512} &
        \qualResult{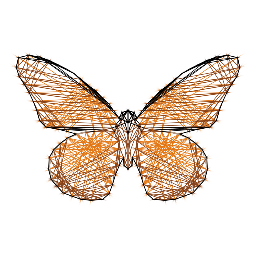}{510}{} &
        \qualResult{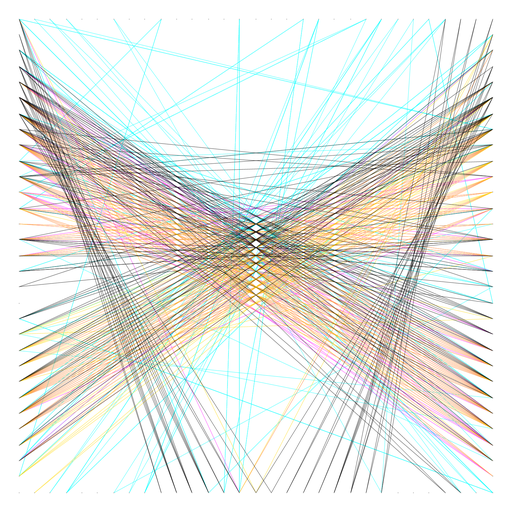}{}{} &
        \qualResult{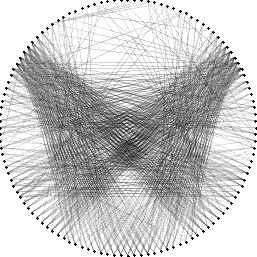}{}{} &
        \qualResult{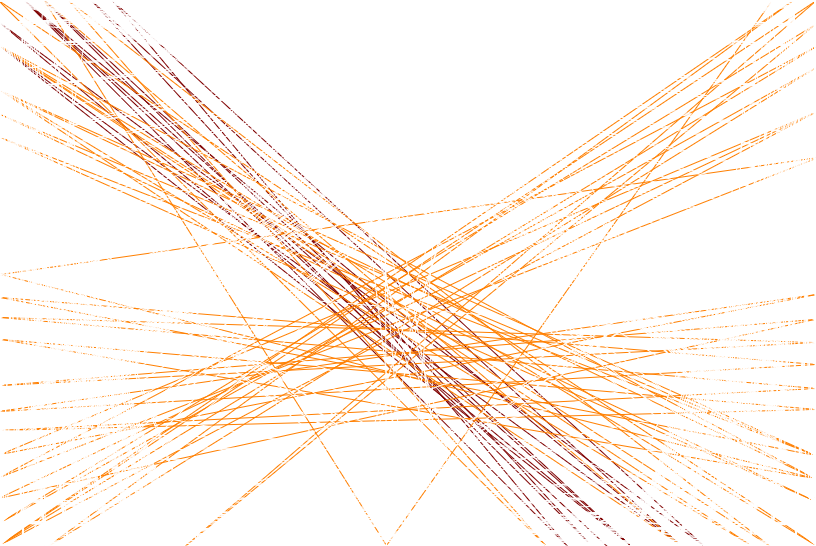}{}{} \\[\qualRowGap]
        \qualBudget{1024} &
        \qualResult{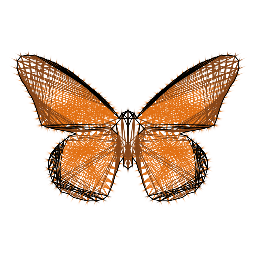}{974}{} &
        \qualResult{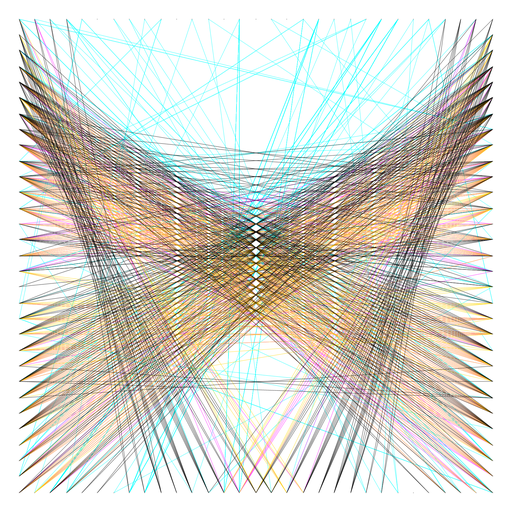}{}{} &
        \qualResult{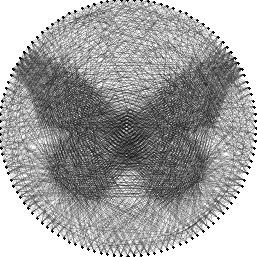}{}{} &
        \qualResult{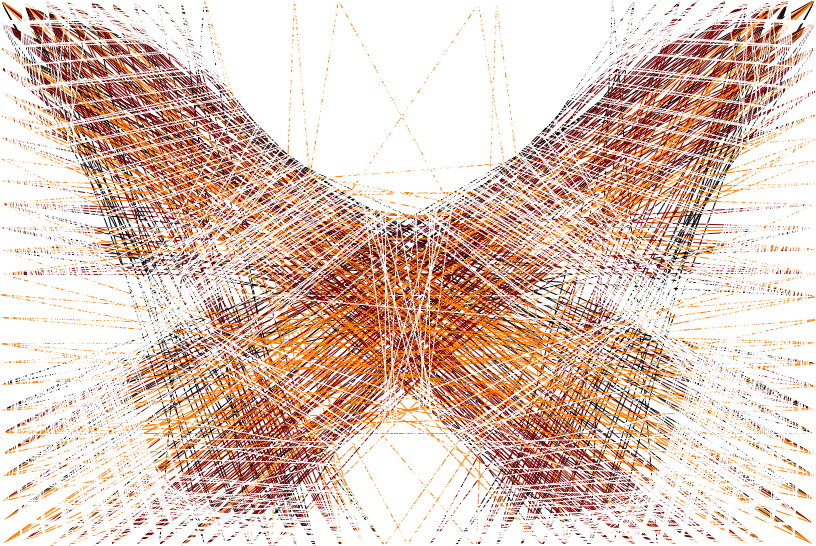}{}{}
      \end{tabular}
    \end{tabular}
  \end{minipage}%
  \hspace{\qualPanelGap}%
  \begin{minipage}[t]{\qualPanelWidth}
    \centering
    \begin{tabular}{@{}c@{\hspace{\qualCellGap}}c@{}}
      \qualTarget{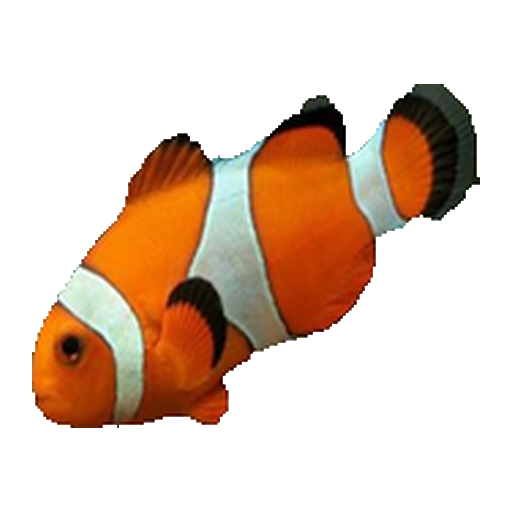}{FSS-1000} &
      \begin{tabular}{@{}c@{\hspace{\qualCellGap}}*{3}{c@{\hspace{\qualCellGap}}}c@{}}
        & \textbf{Ours} & DanAla & SpiderPrinter & Roy (color) \\[2pt]
        \qualBudget{128} &
        \qualResult{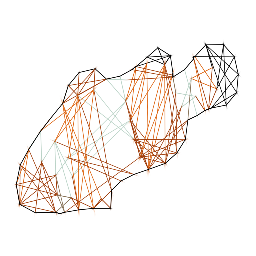}{135}{} &
        \qualResult{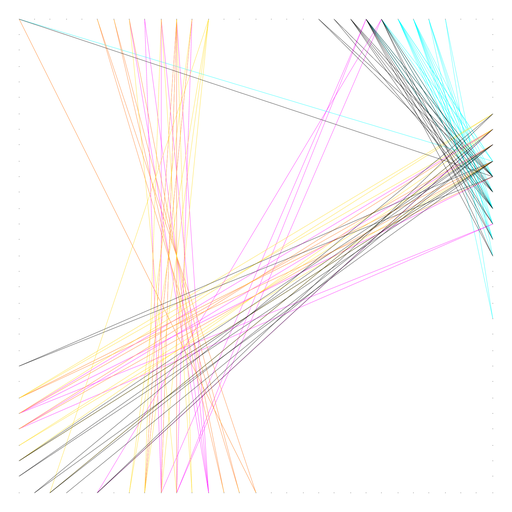}{}{} &
        \qualResult{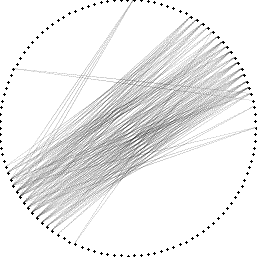}{}{} &
        \qualResult{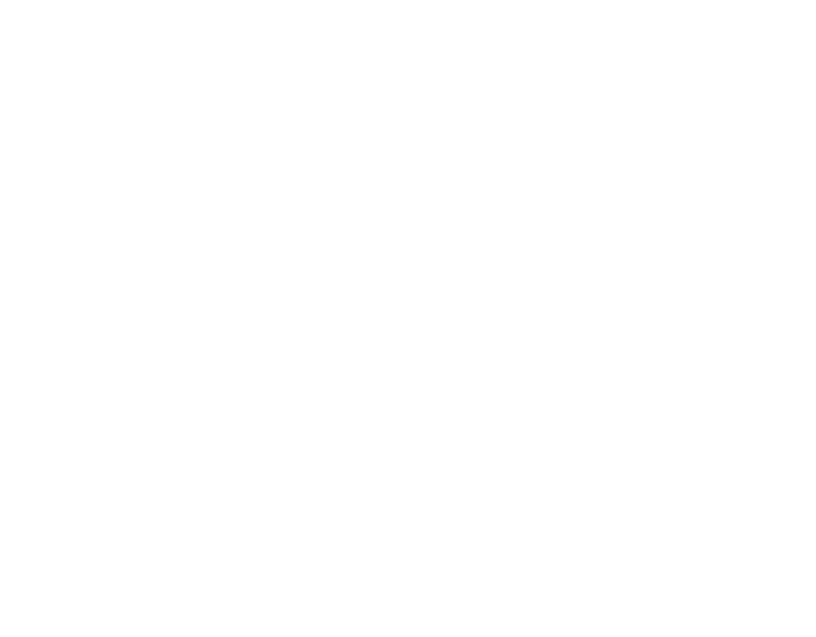}{}{\dagger} \\[\qualRowGap]
        \qualBudget{256} &
        \qualResult{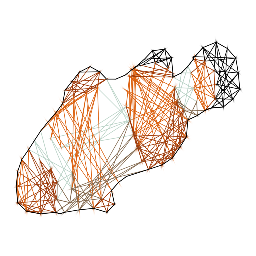}{275}{} &
        \qualResult{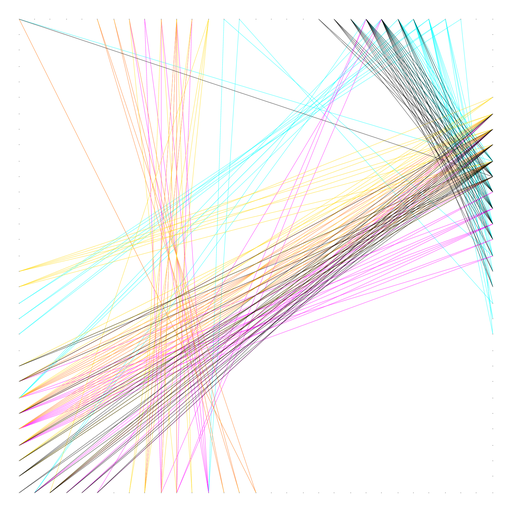}{}{} &
        \qualResult{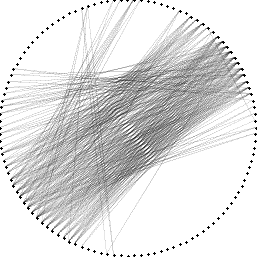}{}{} &
        \qualResult{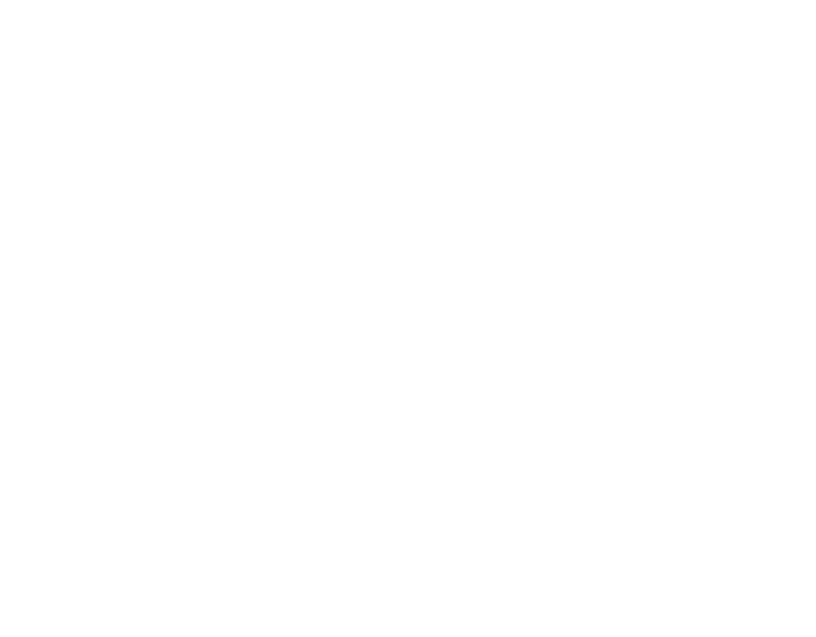}{}{\dagger} \\[\qualRowGap]
        \qualBudget{512} &
        \qualResult{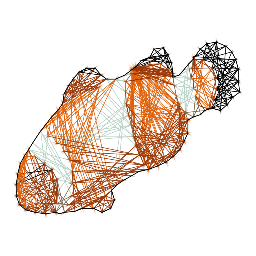}{534}{} &
        \qualResult{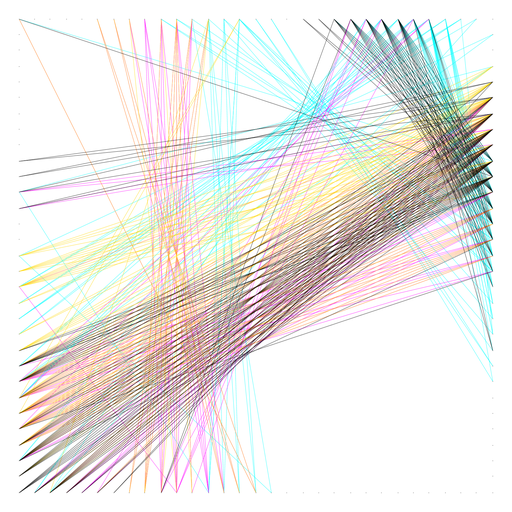}{}{} &
        \qualResult{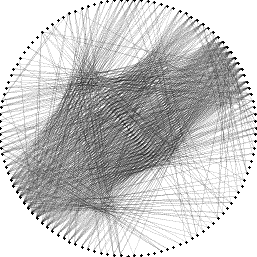}{}{} &
        \qualResult{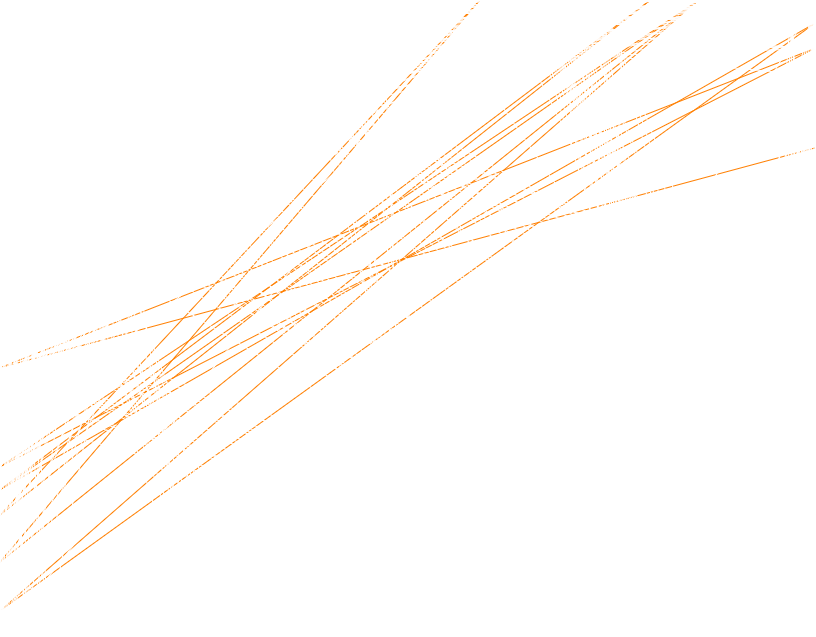}{}{} \\[\qualRowGap]
        \qualBudget{1024} &
        \qualResult{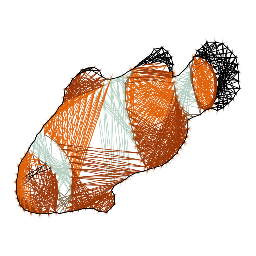}{1019}{} &
        \qualResult{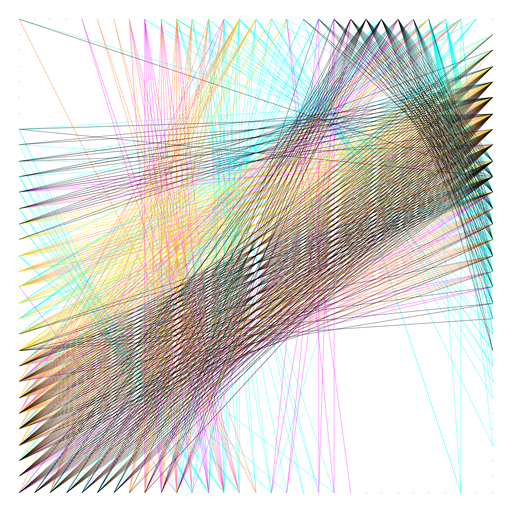}{}{} &
        \qualResult{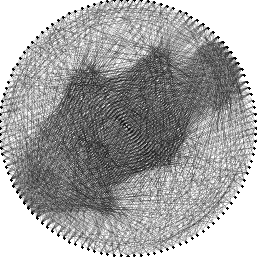}{}{} &
        \qualResult{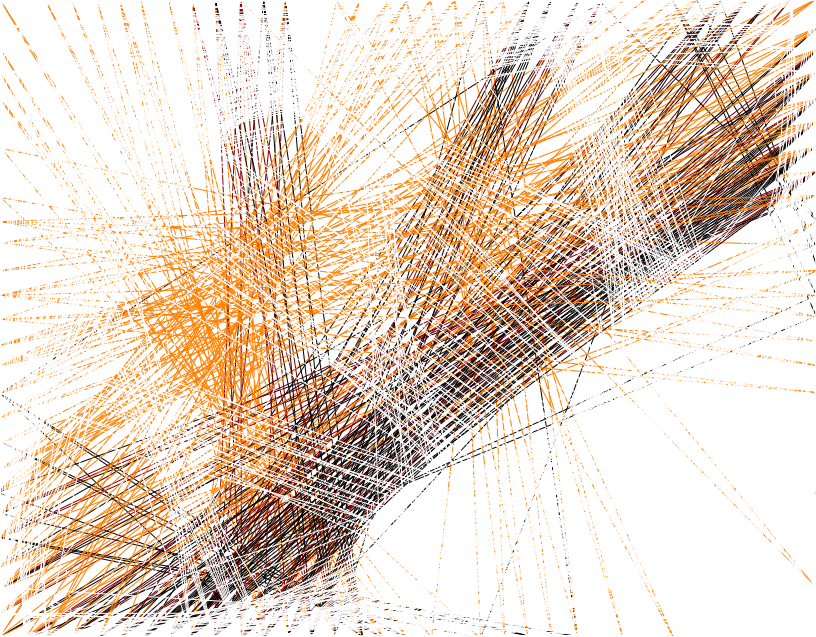}{}{}
      \end{tabular}
    \end{tabular}
  \end{minipage}%
  }%
  \endgroup
  \caption{\textbf{Qualitative comparison at 4 expected string budgets.} Each row corresponds to a budget $B$, and only our results report the actual chord count $K$. Our method is the only one that supports freeform boundaries, and $\dagger$ marks all-white outputs from Roy’s method.}
  \label{fig:qualitative-comparison}
\end{figure*}

%% file: sec/05_experiments.tex

\input{figures/tab_quantitative_comparison}

\section{Experiments}
\label{sec:experiments}

We evaluate image fidelity, the perceived organization of string segments, computational cost, and physical fabrication. We report quantitative RGB comparisons on Butterfly~\cite{wang2009butterfly} and FSS~\cite{li2020fss} with many other types of objects, qualitative comparisons with state-of-the-art methods, ablation studies, performance, physical fabrications, and a user study.

\subsection{Comparison with State-of-the-Art Methods}
\label{sec:sota-comparison}

We compare with three open-source, image-based string-art methods supporting specified or approximately controllable string counts: DanAla's String Art method~\cite{danalaStringArt}, Roy Hachnochi's Multicolor method~\cite{hachnochiStringArt}, and SpiderPrinter~\cite{je2019}. We adopt their original implementations with default settings except the string-count budget. All methods are evaluated at four budgets \(B \in \{128,256,512,1024\}\). We use 180 images from the Leeds Butterfly Dataset~\cite{wang2009butterfly} and 250 images from the FSS-1000 Dataset~\cite{li2020fss}. Using the mask annotations provided with each dataset, we extract the foreground object and set the background to white; all processed targets are resized to $256\times256$. Images from the Leeds Butterfly Dataset are further transformed by rotation, translation, and mirroring so that each butterfly is strictly left-right symmetric. The baselines produce exactly $B$ string segments, whereas our method uses a soft budget loss on string segments to approach $B$, allowing small increases or decreases in the final string count for better visual fidelity. 
Full comparison results are provided in the supplementary material.

\paragraph{Qualitative comparison.}
Fig.~\ref{fig:qualitative-comparison} presents qualitative comparisons on one representative sample from Butterfly and one from FSS. At low budgets, Roy’s method with its original default configuration selects exclusively white strings on the white background, yielding all-white renderings in 96.85\%, 87.67\% at \(B=128,256\), respectively. We also experimented with minor modifications to Roy’s algorithm, but the modified variant still did not yield reasonable renderings. The original outputs are therefore retained in the evaluation and marked with $\dagger$. Our method produces more structured string arrangements, in which near-parallel and envelope-like families of string segments are more frequently observed, whereas the baselines yield visually more cluttered segment distributions. In addition, our results remain well aligned with the target image at low string counts, whereas the baselines struggle to maintain comparable alignment.

\paragraph{Quantitative comparison.}
Table~\ref{tab:quantitative-comparison} reports LPIPS~\cite{zhang2018lpips} and global DINO distance~\cite{oquab2024dinov2}. For each budget, metrics are averaged over the same images, and Mean equally averages the four budget means. Because SpiderPrinter outputs grayscale renderings, its metrics are computed against the grayscale-converted target; all other methods output color renderings and are evaluated directly against the color target. Our method achieves the lowest LPIPS and global DINO distance at every budget and in the four-budget mean on both cohorts, with mean LPIPS/DINO of 0.319/0.558 on Butterfly and 0.369/0.742 on FSS.

\subsection{Ablation Studies}
\label{sec:ablations}
We perform five ablation comparisons: the full model and four variants, namely \textit{greedy method}, \textit{without generatrix}, \textit{without seam handling}, and \textit{without generatrix re-search}. All ablations share the same initialization: region partitions, pin locations, candidate chord sets, and initial color settings. The \textit{greedy method} follows classic string art generation algorithms~\cite{birsak2018,demoussel2022,je2019,oleti2026,schouten2022}, selecting chords greedily instead of optimizing via our differentiable vector renderer. \textit{Without generatrix} removes generatrix parameterization while keeping the differentiable vector renderer. \textit{Without seam handling} disables seam treatment based on M\"obius-manifold. \textit{Without generatrix re-search} retains generatrix gradient refinement and dynamic color selection, but turns off re-search. We further ablate the loss objective terms, whose configurations and results are given in Appendix~D.

Fig.~\ref{fig:ablation-qualitative} presents qualitative comparisons on an FSS sample (a) with budget $B=512$. More results for the four variants are provided in the supplementary material. \textit{Greedy method} (b) produces incorrectly colored black strings with disordered layouts. The variant \textit{Without generatrix} (c) reduces miscolored strings yet yields chaotic structures. \textit{Without seam handling} (d), the overall structure remains regular, while string transitions are abrupt at the upper-right seam region. \textit{Without generatrix re-search} (e), the layout is coherent but contains undesirable strings, such as the blue artifacts on the left. The full model (f) is free of miscolored and spurious strings, achieving regular structures with smooth transitions.

These results verify the necessity of each component: the differentiable vector renderer enables global coordination of color and layout to avoid miscoloration caused by greedy strategies; generatrix parameterization facilitates coherent and dense string arrangements; seam handling guarantees continuous transitions across periodic boundaries; generatrix re-search eliminates undesirable strings trapped in local optima. Beyond the qualitative example, we evaluate the perceptual effect of the generatrix representation through a controlled user study in Sec.~\ref{sec:user-study} and Fig.~\ref{fig:user-study-results}.

\input{figures/fig_ablation_qualitative}

\subsection{Performance}
\label{sec:computational-cost}

We profile Butterfly and FSS on an Intel Core i9-9900K CPU and an NVIDIA GeForce RTX 5090 D v2 GPU using PyTorch 2.11.0/CUDA 12.8 under the default configuration. Table~\ref{tab:computational-cost} reports the average initialization and optimization time and peak memory per run for all string budgets. The initialization stage exhibits higher peak memory because it loads and runs SAM~2~\cite{ravi2024sam2} and DINOv2~\cite{oquab2024dinov2}.

\input{figures/tab_computational_cost}

\subsection{Physical Fabrication}
\label{sec:physical-fabrication}
Our numbered templates, winding manuals, stepwise web guidance, and material estimates primarily follow the fabrication-support workflow in the Iwate University thesis~\cite{chuluunbat2026}. Inspired by Birsak et al.'s graph-based path organization~\cite{birsak2018}, we use a Hierholzer-based decomposition into continuous paths within each region and color layer, allowing cuts and restarts without physical transfer segments. Fig.~\ref{fig:physical-fabrication} and Fig.~\ref{fig:teaser-0010011} show 4 handcrafted artworks using our specification generated by M\"obiuString.

\input{figures/fig_physical_fabrication}

\subsection{User Study}
\label{sec:user-study}

We conducted a pairwise A/B preference study with 24 volunteers at a string budget of $B=512$. Each comparison set contained five targets from Butterfly and five targets from FSS. In each trial, anonymized outputs were presented together with the target image, and participants could choose a no-clear-choice option. Participants evaluated each pair on three criteria: organization (O), i.e., the regularity and structured organization of string segments; target resemblance (R), i.e., fidelity to the target image; and aesthetics (A), i.e., overall visual appeal. Each participant evaluated all 70 comparisons (10 targets × 7 method pairs) once on all three criteria. Trial order and the left--right position of methods were randomized independently per participant.

Fig.~\ref{fig:user-study-results} reports the preference. M\"obiuString is preferred over all three baselines under O, R, and A. For the organization criterion in the ablation, 59.6\% of responses favor the full model; 30.8\% report no clear choice, and 9.6\% favor \textit{Without generatrix}. These results indicate that generatrices are important for producing structured string segments.

\input{figures/fig_user_study_results}

\subsection{Discussion}
\label{sec:regularity-limitations}
We restrict optimization to straight generatrices in periodic pin-index coordinates, despite the broader catalog in Appendix~A, to curb search dimensionality and local-optima risk. Future work will increase expressiveness through adaptive nonlinear generatrices and progressive refinement.

Semantic and color cues may yield unsuitable partitions for nearly uniform-colored furry animals or densely patterned objects. Partitioning aware of geometric capacity and appearance reconstruction remains an open direction.

We explored organization measures inspired by fiber-bundle coherence~\cite{portegies2015fbc}, texture regularity~\cite{sun2021regularity}, good continuation and Gestalt grouping~\cite{lezama2017continuation,rajaei2018gestalt}, and displacement structure~\cite{kailath1979displacement,thomas2018displacement}. Our adapted candidates remained sensitive to density, scale, grouping, and pin representation. We therefore assess perceived organization through qualitative comparisons (Sec.~\ref{sec:sota-comparison}) and the user study (Sec.~\ref{sec:user-study}). Future work will develop interpretable, method-independent measures validated against human judgments and robust to changes in segment count, sampling, and boundary layout.

%% file: figures/tab_quantitative_comparison.tex
\begin{table*}[t]
  \centering
  \caption{\textbf{Quantitative comparison at 4 target budgets for string segments.} Mean averages 4 budgets equally. Best values are bold.}
  \label{tab:quantitative-comparison}
  \small
  \setlength{\tabcolsep}{1.3pt}
  \begin{tabular*}{\textwidth}{@{\extracolsep{\fill}}ll*{10}{c}@{}}
    \toprule
    \multirow{2}{*}{Datasets} & \multirow{2}{*}{Methods} & \multicolumn{2}{c}{$B=128$}
        & \multicolumn{2}{c}{$B=256$}
        & \multicolumn{2}{c}{$B=512$}
        & \multicolumn{2}{c}{$B=1024$}
        & \multicolumn{2}{c}{Mean} \\
    \cmidrule(lr){3-4}\cmidrule(lr){5-6}\cmidrule(lr){7-8}\cmidrule(lr){9-10}\cmidrule(lr){11-12}
     & & LPIPS$\downarrow$ & DINO$\downarrow$
      & LPIPS$\downarrow$ & DINO$\downarrow$
      & LPIPS$\downarrow$ & DINO$\downarrow$
      & LPIPS$\downarrow$ & DINO$\downarrow$
      & LPIPS$\downarrow$ & DINO$\downarrow$ \\
    \midrule
    \multirow{4}{*}{Butterfly} & Ours
      & \textbf{0.372} & \textbf{0.726} & \textbf{0.339} & \textbf{0.611} & \textbf{0.310} & \textbf{0.488} & \textbf{0.256} & \textbf{0.406} & \textbf{0.319} & \textbf{0.558} \\
    & DanAla
      & 0.624 & 0.923 & 0.626 & 0.890 & 0.633 & 0.861 & 0.656 & 0.844 & 0.635 & 0.879 \\
    & SpiderPrinter
      & 0.721 & 0.905 & 0.720 & 0.910 & 0.745 & 0.898 & 0.823 & 0.844 & 0.752 & 0.889 \\
    & Roy Multicolor
      & 0.626 & 0.964 & 0.603 & 0.960 & 0.481 & 0.923 & 0.404 & 0.852 & 0.529 & 0.925 \\
    \midrule
    \multirow{4}{*}{FSS} & Ours
      & \textbf{0.431} & \textbf{0.823} & \textbf{0.393} & \textbf{0.765} & \textbf{0.349} & \textbf{0.711} & \textbf{0.300} & \textbf{0.668} & \textbf{0.369} & \textbf{0.742} \\
    & DanAla
      & 0.614 & 0.941 & 0.612 & 0.934 & 0.621 & 0.929 & 0.641 & 0.929 & 0.622 & 0.933 \\
    & SpiderPrinter
      & 0.705 & 0.927 & 0.699 & 0.926 & 0.717 & 0.926 & 0.781 & 0.909 & 0.726 & 0.922 \\
    & Roy Multicolor
      & 0.673 & 0.948 & 0.664 & 0.947 & 0.597 & 0.940 & 0.505 & 0.897 & 0.610 & 0.933 \\
    \bottomrule
  \end{tabular*}
  \end{table*}

%% file: figures/fig_ablation_qualitative.tex
\begin{figure}[t]
  \centering
  \includegraphics[width=0.96\columnwidth]{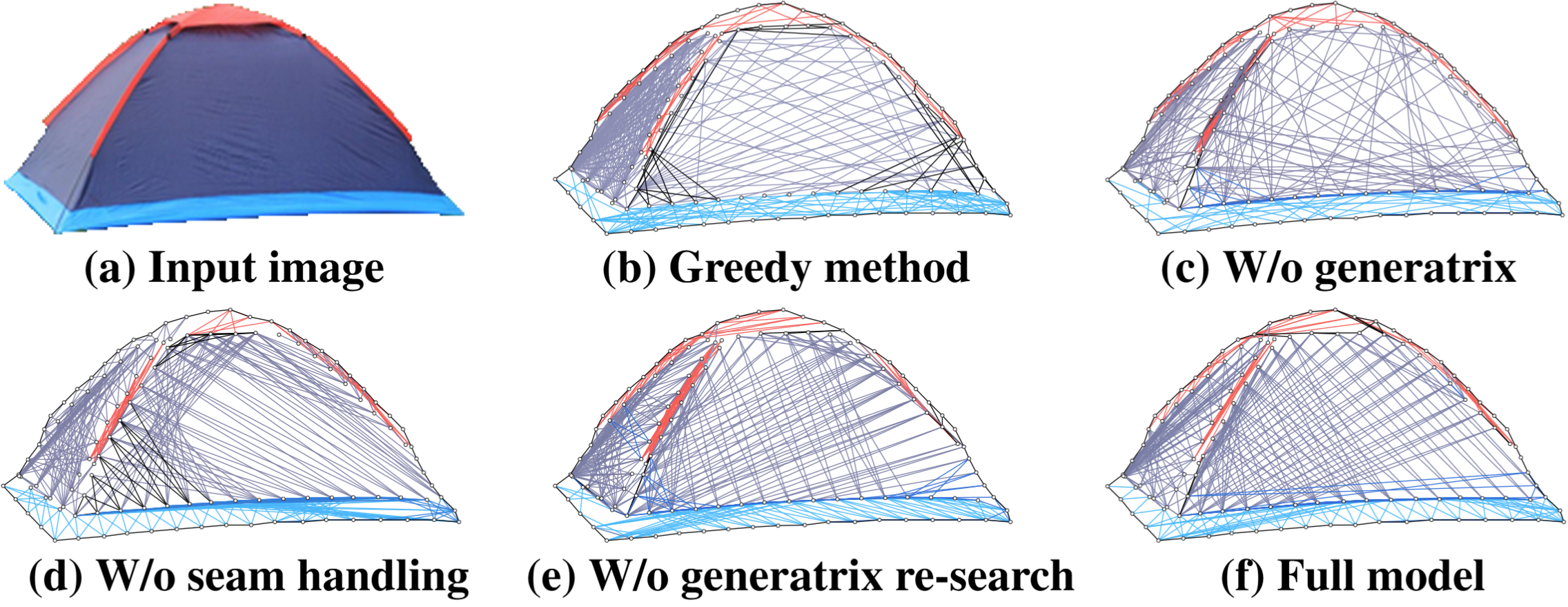}

  \caption{\textbf{Generatrix ablation on FSS at $B=512$.}}
  \label{fig:ablation-qualitative}
\end{figure}

%% file: figures/tab_computational_cost.tex
\begin{table}[t]
  \centering
  \caption{\textbf{Computational costs for two stages.}}
  \label{tab:computational-cost}
  \small
  \setlength{\tabcolsep}{2pt}
  \begin{tabular*}{\columnwidth}{@{\extracolsep{\fill}}lrrrr@{}}
    \toprule
    \multirow{2}{*}{Stages}  & \multicolumn{2}{c}{Butterfly} & \multicolumn{2}{c}{FSS} \\
    \cmidrule(lr){2-3}\cmidrule(l){4-5}
    & Time (s) & GPU (GiB) & Time (s) & GPU (GiB) \\
    \midrule
    Initialization & 171.6 & 1.843 & 163.1 & 1.850 \\
    Optimization & 443.1 & 0.590 & 668.2 & 0.802 \\
    \bottomrule
  \end{tabular*}
\end{table}

%% file: figures/fig_physical_fabrication.tex
\begin{figure}[t!]
  \centering
  \includegraphics[width=1\columnwidth]{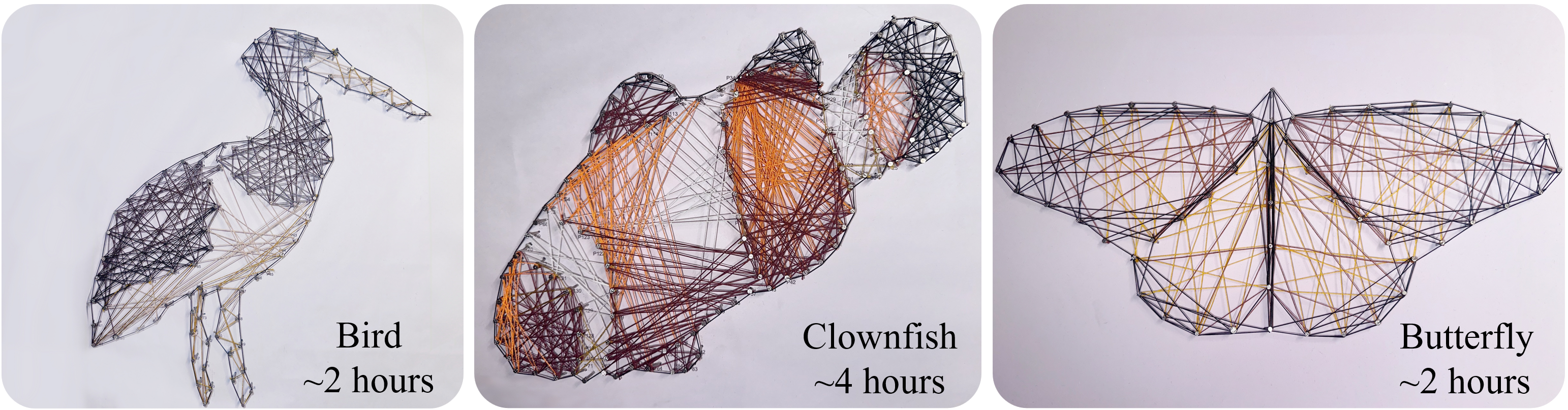}
  \caption{\textbf{Three handcrafted artworks with fabrication time.}}
  \label{fig:physical-fabrication}
\end{figure}

%% file: figures/fig_user_study_results.tex
\begin{figure}[!t]
  \centering
  \input{figures/user_study_plot}
  \caption{\textbf{Statistics of the user study.}
  O/R/A represent organization (regularity), target resemblance, and aesthetics, while the blue/gray/orange in each bar means the first method / no clear choice / second method.}
  \label{fig:user-study-results}
\end{figure}
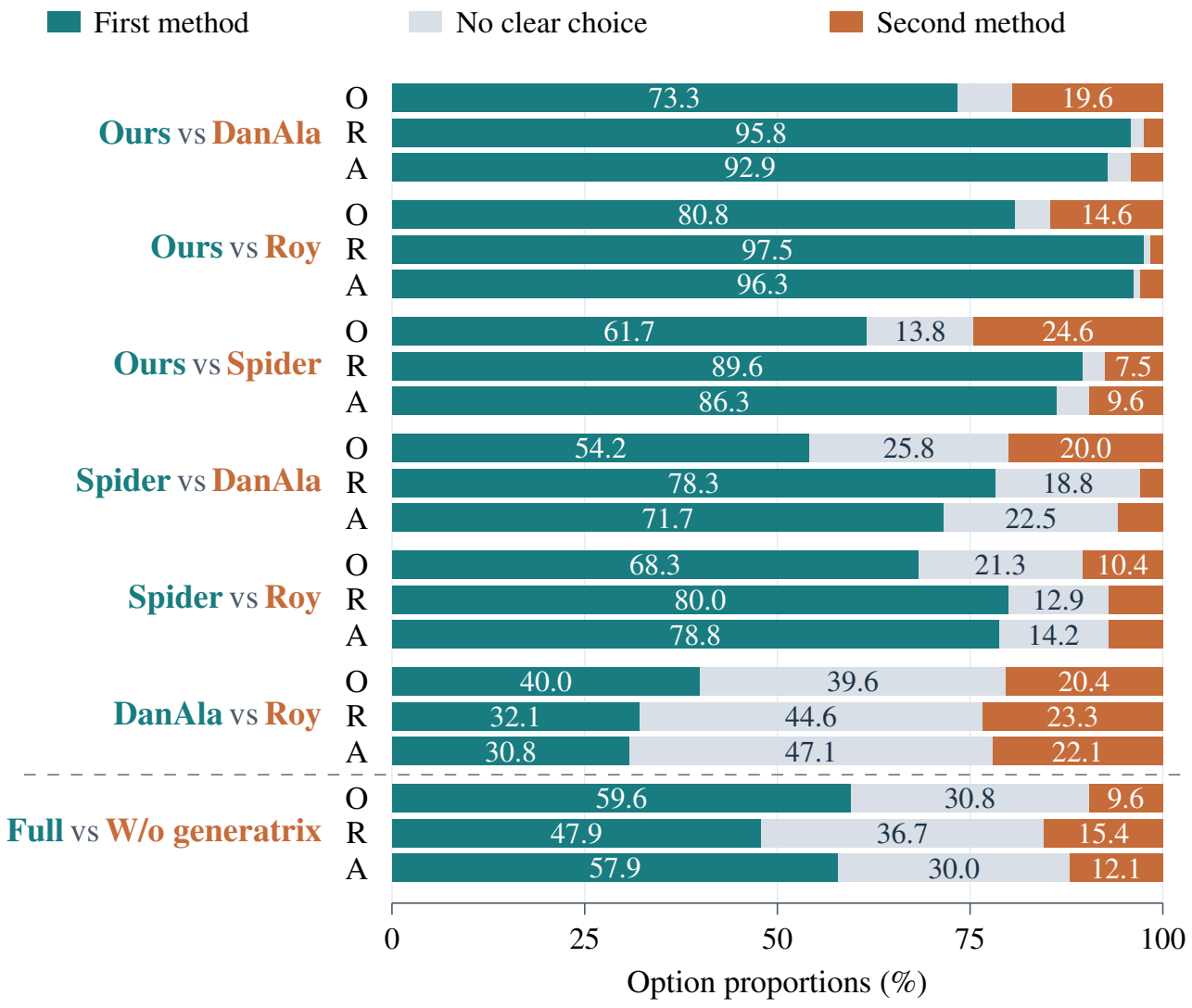

%% file: figures/user_study_plot.tex
\begingroup
\fontfamily{ptm}\selectfont
\setlength{\unitlength}{\dimexpr\linewidth/100\relax}
\definecolor{USFirst}{HTML}{187C80}
\definecolor{USNeutral}{HTML}{D9DFE6}
\definecolor{USSecond}{HTML}{C56C3A}
\definecolor{USInk}{HTML}{1D3348}
\definecolor{USMuted}{HTML}{4B5869}
\definecolor{USGrid}{HTML}{E4E8ED}

\def\USFirstY{90}
\def\USGroupStep{10}
\def\USRowStep{3}
\def\USLabelThreshold{7.5}
\newdimen\USPlotWidth
\newdimen\USPercentUnit
\newdimen\USBarHeight
\newdimen\USFirstWidth
\newdimen\USNeutralWidth
\newdimen\USSecondWidth
\newdimen\USLabelOffset
\newdimen\USRoundDimension
\newcount\USY
\newcount\USRounded
\newcount\USWhole
\newcount\USDecimal
\USPlotWidth=66\unitlength
\USPercentUnit=\dimexpr\USPlotWidth/100\relax
\USBarHeight=2.45\unitlength

\def\USPct#1{%
  \begingroup
  \USRoundDimension=#1pt\relax
  \USRounded=\USRoundDimension
  \USRounded=\numexpr\USRounded*10/65536\relax
  \USWhole=\USRounded \divide\USWhole by10
  \USDecimal=\numexpr\USRounded-10*\USWhole\relax
  \number\USWhole.\number\USDecimal
  \endgroup
}

\def\USPair#1#2#3{%
  \USY=\numexpr\USFirstY-\USGroupStep*#1-\USRowStep\relax
  \put(0,\number\USY){\makebox(25.5,0)[r]{%
    \fontsize{6.8}{8}\selectfont
    \textbf{\textcolor{USFirst}{#2}}\,\textcolor{USMuted}{vs}\,\textbf{\textcolor{USSecond}{#3}}}}%
}

\def\USRow#1#2#3#4#5#6{%
  \USY=\numexpr\USFirstY-\USGroupStep*#1-\USRowStep*#2\relax
  \USFirstWidth=#4\USPercentUnit
  \USNeutralWidth=#5\USPercentUnit
  \USSecondWidth=#6\USPercentUnit
  \put(31.5,\number\USY){\raisebox{-.5\USBarHeight}{\makebox[0pt][l]{%
    \textcolor{USFirst}{\rule{\USFirstWidth}{\USBarHeight}}%
    \textcolor{USNeutral}{\rule{\USNeutralWidth}{\USBarHeight}}%
    \textcolor{USSecond}{\rule{\USSecondWidth}{\USBarHeight}}}}}%
  \put(28.5,\number\USY){\makebox(0,0){\fontsize{6.5}{7}\selectfont #3}}%
  \ifdim#4pt<\USLabelThreshold pt\relax\else
    \USLabelOffset=.5\USFirstWidth
    \put(31.5,\number\USY){\hspace*{\USLabelOffset}\makebox(0,0){%
      \fontsize{6}{7}\selectfont\textcolor{white}{\USPct{#4}}}}%
  \fi
  \ifdim#5pt<\USLabelThreshold pt\relax\else
    \USLabelOffset=\dimexpr\USFirstWidth+.5\USNeutralWidth\relax
    \put(31.5,\number\USY){\hspace*{\USLabelOffset}\makebox(0,0){%
      \fontsize{6}{7}\selectfont\textcolor{USInk}{\USPct{#5}}}}%
  \fi
  \ifdim#6pt<\USLabelThreshold pt\relax\else
    \USLabelOffset=\dimexpr\USFirstWidth+\USNeutralWidth+.5\USSecondWidth\relax
    \put(31.5,\number\USY){\hspace*{\USLabelOffset}\makebox(0,0){%
      \fontsize{6}{7}\selectfont\textcolor{white}{\USPct{#6}}}}%
  \fi
}

\begin{picture}(100,88)(0,11)
  \put(2,96.5){\makebox(0,0)[l]{\textcolor{USFirst}{\rule{2.8\unitlength}{1.8\unitlength}}}}
  \put(6,96.5){\makebox(0,0)[l]{\fontsize{6.2}{7}\selectfont First method}}
  \put(33,96.5){\makebox(0,0)[l]{\textcolor{USNeutral}{\rule{2.8\unitlength}{1.8\unitlength}}}}
  \put(37,96.5){\makebox(0,0)[l]{\fontsize{6.2}{7}\selectfont No clear choice}}
  \put(69,96.5){\makebox(0,0)[l]{\textcolor{USSecond}{\rule{2.8\unitlength}{1.8\unitlength}}}}
  \put(73,96.5){\makebox(0,0)[l]{\fontsize{6.2}{7}\selectfont Second method}}

  {\color{USGrid}\linethickness{0.25pt}
    \multiput(31.5,21)(16.5,0){5}{\line(0,1){70.3}}}%
  {\color{USMuted}\linethickness{0.3pt}
    \put(31.5,21){\line(1,0){66}}
    \multiput(31.5,21)(16.5,0){5}{\line(0,-1){1}}}%
  \fontsize{6.2}{7}\selectfont
  \put(31.5,18){\makebox(0,0){0}}
  \put(48,18){\makebox(0,0){25}}
  \put(64.5,18){\makebox(0,0){50}}
  \put(81,18){\makebox(0,0){75}}
  \put(97.5,18){\makebox(0,0){100}}
  
  \put(64.5,14){\makebox(0,0){\fontsize{6.5}{8}\selectfont Option proportions (\%)}}

  \USPair{0}{Ours}{DanAla}
  \USRow{0}{0}{O}{73.333333333333}{7.083333333333}{19.583333333333}
  \USRow{0}{1}{R}{95.833333333333}{1.666666666667}{2.500000000000}
  \USRow{0}{2}{A}{92.916666666667}{2.916666666667}{4.166666666667}

  \USPair{1}{Ours}{Roy}
  \USRow{1}{0}{O}{80.833333333333}{4.583333333333}{14.583333333333}
  \USRow{1}{1}{R}{97.500000000000}{0.833333333333}{1.666666666667}
  \USRow{1}{2}{A}{96.250000000000}{0.833333333333}{2.916666666667}

  \USPair{2}{Ours}{Spider}
  \USRow{2}{0}{O}{61.666666666667}{13.750000000000}{24.583333333333}
  \USRow{2}{1}{R}{89.583333333333}{2.916666666667}{7.500000000000}
  \USRow{2}{2}{A}{86.250000000000}{4.166666666667}{9.583333333333}

  \USPair{3}{Spider}{DanAla}
  \USRow{3}{0}{O}{54.166666666667}{25.833333333333}{20.000000000000}
  \USRow{3}{1}{R}{78.333333333333}{18.750000000000}{2.916666666667}
  \USRow{3}{2}{A}{71.666666666667}{22.500000000000}{5.833333333333}

  \USPair{4}{Spider}{Roy}
  \USRow{4}{0}{O}{68.333333333333}{21.250000000000}{10.416666666667}
  \USRow{4}{1}{R}{80.000000000000}{12.916666666667}{7.083333333333}
  \USRow{4}{2}{A}{78.750000000000}{14.166666666667}{7.083333333333}
  
  \USPair{5}{DanAla}{Roy}
  \USRow{5}{0}{O}{40.000000000000}{39.583333333333}{20.416666666667}
  \USRow{5}{1}{R}{32.083333333333}{44.583333333333}{23.333333333333}
  \USRow{5}{2}{A}{30.833333333333}{47.083333333333}{22.083333333333}

  \USPair{6}{Full}{W/o generatrix}
  \USRow{6}{0}{O}{59.583333333333}{30.833333333333}{9.583333333333}
  \USRow{6}{1}{R}{47.916666666667}{36.666666666667}{15.416666666667}
  \USRow{6}{2}{A}{57.916666666667}{30.000000000000}{12.083333333333}

  {\color{USMuted}\linethickness{0.25pt}
    \multiput(0,32)(2,0){50}{\line(1,0){1}}}%

\end{picture}%
\endgroup

%% file: sec/06_discussion.tex
\section{Conclusion}
\label{sec:conclusion}

We present \textit{M\"obiuString}, a string art generation framework with structured string organization on freeform, image-derived region boundaries. It introduces the generatrix on the M\"obius strip of non-oriented boundary-point pairs to formalize structured families of string segments, and jointly optimizes the parameters of all generatrices. Given an image and a string budget, pins are placed along boundaries of image-derived regions, and candidate string segments near each generatrix in the homeomorphic pin-index plane serve as initialization. Multi-color strings on multiple layers are jointly optimized through differentiable vector rendering based on 2D Gaussian Splatting, with periodic discrete generatrix re-search to avoid local optima. Evaluations with two datasets, ablation studies, and a user study demonstrate the contribution of each proposed component and superior performance against baselines.

%% file: sec/A_proofs.tex
\section{Generatrices and Envelopes in Chord Space}
\label{app:proofs}
\numberwithin{equation}{subsection}

To explain the patterns in Fig.~2 of the main paper, we derive the envelope of the lines joining the moving endpoints of a family of string segments.

\subsection{Unordered endpoints and the M\"obius seam}
\label{app:topology}
Let $\gamma:S^1\to\partial\Omega$ parameterize a simple closed boundary once. Angles $\alpha,\beta$ label positions along it, including on freeform regions. Swapping them preserves the segment. Including coincident pairs gives the chord space $\mathcal M=\operatorname{Sym}^2(S^1)$ of Eq.~(1) in the main paper~\cite{morton1967}.

Choose real lifts of the two angles and put
\begin{equation}
 \theta=(\alpha+\beta)/2,\qquad r=\cos((\beta-\alpha)/2).
 \label{eq:topologymap}
\end{equation}
Exchanging the endpoints preserves $(\theta,r)$. Adding $2\pi$ to either angle sends it to $(\theta+\pi,-r)$, so the vertical edges of $[0,\pi]\times[-1,1]$ are glued with reversed height. The inverse recovers the unordered pair $\{\theta-\arccos r,\theta+\arccos r\}$ modulo $2\pi$. These maps identify $\mathcal M$ with the M\"obius band. Its single boundary, $r=\pm1$, consists of coincident pairs and is excluded from rendering.

The same seam appears in the triangular diagram of Sec.~3.2. Number the $n$ pins around a region and sort each pair so that $0\le i\le j\le n$. Pin $n$ is a second copy of pin $0$, so $(i,n)$ continues at $(0,i)$. The two marks describe the same physical segment. The diagonal $i=j$, in contrast, consists of zero-length segments. Finite pins sample this continuous space, and a concave region further restricts the sample to segments lying inside the region.

\subsection{Hough coordinates of a string segment}
\label{app:convexhough}
Fix an image origin $o$. Write $\normal(\phi)=(\cos\phi,\sin\phi)$ and $\tangent(\phi)=(-\sin\phi,\cos\phi)$. A line with unit normal $\normal$ and signed offset $\rho$ satisfies
\begin{equation}
 \normal(\phi)\cdot(x-o)=\rho .
 \label{eq:appAline}
\end{equation}
For distinct endpoints $A,B$, the normal is perpendicular to $B-A$. The pairs $(\phi,\rho)$ and $(\phi+\pi,-\rho)$ describe the same line. On a disk of radius $R$ centered at $o$, taking $\alpha,\beta$ as polar angles gives $\phi=\theta\bmod\pi$. Choosing the normal at $\phi=\theta$ and projecting either endpoint onto it gives
\begin{equation}
 \rho=R\cos((\beta-\alpha)/2)=Rr.
 \label{eq:circularhough}
\end{equation}
Here $r=0$ represents diameters. The values $|\rho|=R$ describe the limiting tangents at coincident endpoints.

For a smooth strictly convex region, let $H(\phi)=\max_{x\in\overline\Omega}\normal(\phi)\cdot(x-o)$. A line cuts the interior exactly when
\begin{equation}
 -H(\phi+\pi)<\rho<H(\phi).
 \label{eq:admissibleoffset}
\end{equation}
The two boundary intersections recover the segment. On a concave region a line may cut several disjoint intervals, and on a flat boundary several pairs may lie on the same line. Endpoint indices retain the identity of the selected pair in both cases.

\subsection{From a generatrix to an image-space envelope}
\label{app:lineenvelope}
\input{figures/fig_analytic_generatrices}
On a smooth branch $\rho=h(\phi)$, write $F(x,\phi)=\normal(\phi)\cdot(x-o)-h(\phi)$. The limiting intersection $E$ of neighboring lines satisfies $F(E,\phi)=0$. Subtracting their line equations and taking the limit gives $F_\phi(E,\phi)=0$. Since $\normal'=\tangent$,
\begin{align}
 \normal\cdot(E-o)&=h,\qquad \tangent\cdot(E-o)=h',\nonumber\\
 E(\phi)&=o+h\normal+h'\tangent .
 \label{eq:generalenvelope}
\end{align}
Differentiating yields $E'=(h+h'')\tangent$. Wherever $h+h''\ne0$, the selected line is tangent to the envelope at $E$~\cite{sir2008,quenell2009}. If all the lines pass through one point, $E$ stays at that point. A family of distinct parallel lines has a fixed normal and no finite envelope point.

In a disk, the endpoints are $o+h\normal\pm s\tangent$, with $s=\sqrt{R^2-h^2}$ for $|h|<R$. The tangency point has tangential coordinate $h'$, so it lies on the finite segment exactly when
\begin{equation}
 |h'|\le s \quad\Longleftrightarrow\quad h^2+(h')^2\le R^2 .
 \label{eq:finiteenvelope}
\end{equation}
Equality places contact at an endpoint. A constant offset $h=c$ gives a circle of radius $|c|$, degenerating to its center for $c=0$.

\subsection{Ten analytic generatrices on the unit disk}
\label{app:analytic}
Panel (b) uses reversion through an interior point~\cite{kocik2013}. Panels (c)--(f) and (j) use the support-function construction~\cite{sir2008}, with parameters chosen for illustration.

For the unit disk in Fig.~\ref{fig:analytic-generatrix-atlas}, $o=0$ and the endpoints for a given $h(\phi)$ have angles $\phi\pm\arccos h(\phi)$.

\paragraph{(a) Boundary fan.}
Fix $\beta=3\pi/4$ and let $\alpha$ traverse the circle. All segments share one boundary point, and their envelope collapses to that endpoint.

\paragraph{(b) Interior point bundle.}
Set $h=P\cdot\normal$ over a half-turn of $\phi$, with $P=(0.48,0.15)$. All lines pass through $P$, and Eq.~\eqref{eq:generalenvelope} gives $E=P$.

\paragraph{(c) Eccentric circle.}
Set $h=P\cdot\normal+0.26$ with $P=(0.40,0.15)$. Over one turn, $E=P+0.26\normal$ traces a circle centered at $P$.

\paragraph{(d) Eccentric ellipse.}
Let $Q$ have eigenvalues $a^2,b^2$ along the principal axes of an ellipse centered at $P$. Its tangent lines are specified by
\begin{equation}
 h=P\cdot\normal+\sqrt{\normal^TQ\normal},\qquad
 E=P+\frac{Q\normal}{\sqrt{\normal^TQ\normal}} .
 \label{eq:appAellipse}
\end{equation}

\paragraph{(e) Nested ellipses and a point bundle.}
Scaling the support term $\sqrt{\normal^TQ\normal}$ by $\lambda\in\{0,0.5,1\}$ gives a point bundle at $P$ and tangents to two concentric ellipses.

\paragraph{(f) Periodic convex envelope.}
Set $h=0.53[1+0.105\cos(3(\phi-\pi/12))]$. Its period $2\pi/3$ and positive $h+h''$ give a smooth convex envelope with threefold symmetry.

\paragraph{(g) Cardioid.}
Pair $\alpha$ with $\beta=2\alpha+0.35$. Write $p(\alpha)=(\cos\alpha,\sin\alpha)$. Substituting $\phi=(\alpha+\beta)/2$ and $h=\cos((\beta-\alpha)/2)$ into Eq.~\eqref{eq:generalenvelope} gives $E=(2p(\alpha)+p(\beta))/3$, a cardioid~\cite{bouthillier2018,herr2026}. The cusp occurs when the two endpoints are antipodal. The endpoint slope here is $2$.

\paragraph{(h) Blaschke/Poncelet ellipse.}
Identify the unit circle with $|z|=1$ and choose $a=-0.30+0.10\,\mathrm{i}$ and $b=0.45+0.20\,\mathrm{i}$. Define
\begin{equation}
 B(z)=z\,\frac{z-a}{1-\bar a z}\,\frac{z-b}{1-\bar b z}.
\end{equation}
By the theorem of Daepp, Gorkin, and Mortini, stated in Lemma~2.1 of Fujimura~\cite{fujimura2017}, joining the three preimages of any $|\lambda|=1$ under $B$ gives a triangle whose sides are tangent to $|z-a|+|z-b|=|1-\bar a b|$. The three-step closure distinguishes this construction from the freely chosen ellipse in (d).

\paragraph{(i) Local parabola.}
Dhar et al.~\cite{dhar2025} derive parabolic envelopes in string art. For (i), use image-aligned axes $(u,v)$ centered at $P=(0,-0.35)$ and take the lines $v-qu+cq^2=0$ with $c=0.30$ and $|q|\le1.25$. Differentiating the line equation with respect to $q$ gives $u=2cq$, hence $v=cq^2$. The boundary intersections supply the segment endpoints, and the restricted range of $q$ selects the visible parabolic arc.

\paragraph{(j) Involute from a straight Hough generatrix.}
Take $h(\phi)=0.12\phi+0.10$ for $0\le\phi\le5.6$. The envelope is $E=h\normal+0.12\tangent$. The point $0.12\tangent$ travels on a circle of radius $0.12$, and $h\normal$ lies along its tangent with length increasing by $0.12$ per radian. This is the unwinding construction of a circle involute. Equation~\eqref{eq:finiteenvelope} holds over the plotted interval.

\subsection{The two slopes used by M\"obiuString}
\label{app:chartrelation}
Section~3.2 of the main paper uses finite straight segments in a lifted pin-index plane, with intercept $b_m$, extent $[u_m,v_m]$, and fixed slope $k_m$,
\begin{equation}
 j=k_m i+b_m,\quad u_m\le i\le v_m,\quad k_m=\pm1.
 \label{eq:appAindexslot}
\end{equation}
On $n$ uniformly spaced circular pins, $\alpha=2\pi i/n$ and $\beta=2\pi j/n$. Equations~\eqref{eq:topologymap} and~\eqref{eq:circularhough} then give the coordinate change
\begin{equation}
 \phi=\pi(i+j)/n,\qquad \rho=R\cos\!\bigl(\pi(j-i)/n\bigr).
\end{equation}
It explains the two patterns in Fig.~2 of the main paper.

\paragraph{Slope $+1$ fixes the endpoint separation.}
For $j=i+b_m$, the angular difference is $2\pi b_m/n$. Equation~\eqref{eq:circularhough} gives constant offset $R\cos(\pi b_m/n)$ while the normal rotates. The tangency points trace an arc of a concentric circle, collapsing to the center for diameters.

\paragraph{Slope $-1$ fixes the mean angle.}
For $j=-i+b_m$, the offset varies at fixed normal angle $\pi b_m/n$ modulo $\pi$, giving parallel lines.

The slope in panel (j) is $dh/d\phi$ in Hough coordinates. M\"obiuString uses only the two fixed pin-index slopes and optimizes the placement and extent of each generatrix. On freeform boundaries the resulting pattern also depends on the boundary shape.

%% file: figures/fig_analytic_generatrices.tex
\begin{figure*}[t]
  \centering
  \includegraphics[width=\textwidth]{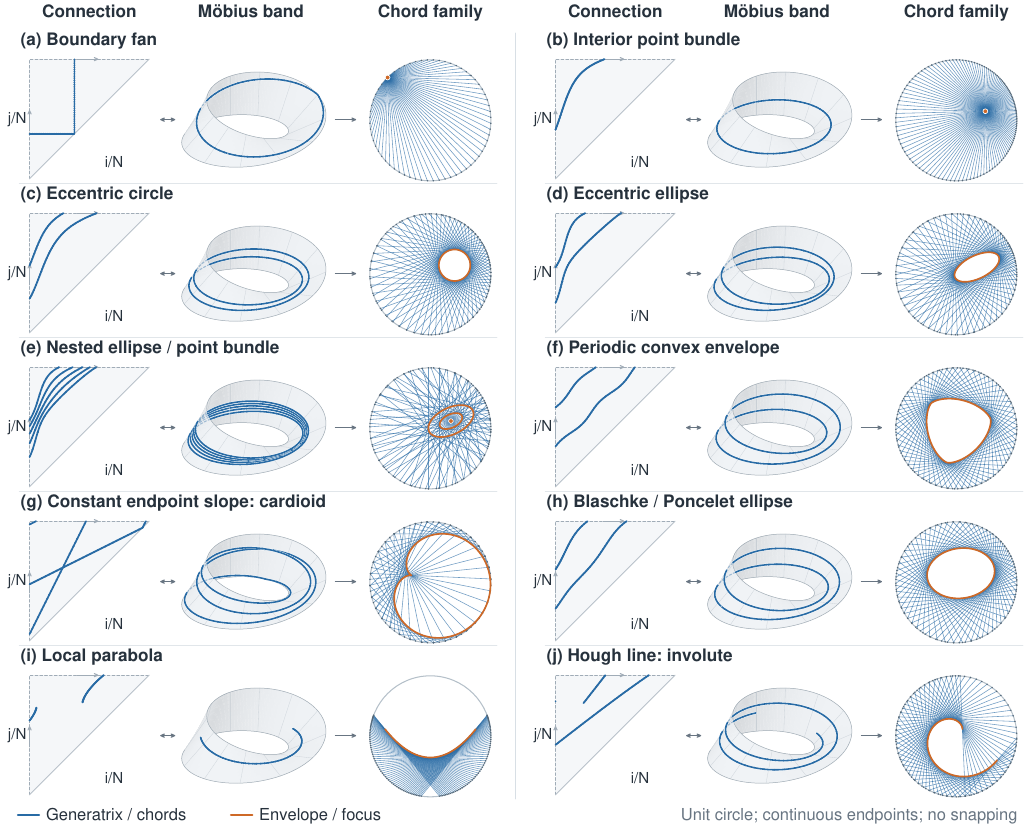}
  \caption{\textbf{Analytic generatrices in three corresponding views.}
  Each panel shows a normalized endpoint-index connection diagram, its M\"obius-band image, and the associated chords in the unit disk, from left to right.
  The three views share the same continuous, unsnapped endpoint pairs.
  Blue denotes generatrices and chords; orange denotes an envelope or degenerate focus; dashed band curves are occluded.
  The connection chart is $0\le i/N\le j/N\le1$, with identified edges $(0,t)\sim(t,1)$.
  Panels (a)--(j) each display $64$ chords; their formulas and parameter choices are given in the supplementary text.
  The index-slope $\pm1$ examples already shown in the main paper are omitted here.
  Envelopes describe tangency boundaries, not filled interiors.
  These are analytic dictionary examples, not additional enabled optimization slots.}
  \label{fig:analytic-generatrix-atlas}
\end{figure*}

%% file: sec/B_budget.tex
\section{Pin Spacing under a String Budget}
\label{app:budget}
\counterwithout{equation}{subsection}
\numberwithin{equation}{section}

\paragraph{From pin spacing to candidate pairs.}
Sampling a region of perimeter $P_r$ at arc-length spacing $h$ gives $n_r\simeq P_r/h$ pins. Each of its $L_r$ color layers has $\binom{n_r}{2}$ unordered pin pairs. Summing over regions gives the candidate count before filtering,
\begin{equation}
 C(h)=\sum_r L_r\,\frac{n_r(n_r-1)}{2}
 \;\simeq\; \frac{1}{2h^2}\sum_r L_r P_r^2 .
 \label{eq:capacity}
\end{equation}
The approximation retains the leading term in $n_r$.

\ifdefined\pinSpacingProofBreak\pinSpacingProofBreak\fi
\paragraph{Why an inverse square root?}
For fixed regions and color layers, $C(h)\propto h^{-2}$ to leading order. Choosing $C(h(B))\propto B$ for target budget $B$ gives
\begin{equation}
 h(B)\propto B^{-1/2}, \qquad
 \frac{h(B_2)}{h(B_1)}=\sqrt{\frac{B_1}{B_2}} .
 \label{eq:budgetspacing}
\end{equation}
The spacing rule in Sec.~4.1 of the main paper uses this scaling. Quadrupling the budget halves the spacing.

%% file: sec/C_implementation.tex
\section{Color Losses and Soft Counting}
\label{app:implementation}
\counterwithout{equation}{subsection}
\numberwithin{equation}{section}

Let $V_k(x)$ and $V_o(x)$ be the visible shares of filling color $c_k$ and the fixed black outline at pixel $x$. Total coverage is $A=V_o+\sum_k V_k$, and premultiplied RGB is $P=\sum_k V_kc_k$. The target color mixture gives the demand $F(x)\in[0,1]$ for nonwhite material. It vanishes on bare background and can be low in a white foreground highlight. For RGB values in $[0,1]^3$, color similarity to the target is
\begin{equation}
 s(c,x)=\exp\!\left(-\frac{\|c-I^\star(x)\|_2^2}{2\sigma_c^2}\right),
 \qquad \sigma_c=0.20 .
 \label{eq:appCsim}
\end{equation}

\paragraph{Mis-color loss.}
The visible mismatch from filling segments is $W=\sum_k V_k[1-s(c_k,x)]$. For a supported target color $k$ in region group $g$, let $\pi_k(x)$ be its share of target material at $x$. The demand for $j=(g,k)$ is $D_j(x)=\ind_g(x)F(x)\pi_k(x)$, where $\ind_g$ indicates group membership. For a Gaussian window $G_\nu$ of width $\nu\in\{2,4,8\}$ pixels, the local mismatch is
\begin{equation}
 w_{j\nu}=\frac{G_\nu*(D_j W)}{G_\nu*D_j}.
\end{equation}
Write $\langle z\rangle_H=\sum_x H(x)z(x)/\sum_x H(x)$. Below, $\mathbb E_H$ averages over pixels with window weights $H_{j\nu}$, then over target colors within each group and over groups with weights proportional to the square root of their respective demand masses. Scales have equal weight. Empty supports are omitted and the remaining weights normalized. With $\rho_0=0.35$, the loss is
\begin{align}
 \mathcal L_{\rm mis}={}&
 \frac{\sqrt{\mathbb E_{G_\nu*D_j}[w_{j\nu}^2]}
       +\langle V_o[1-s(\mathbf0,x)]\rangle_F}{\rho_0}
 \nonumber\\
 &+0.25 E_{\rm RGB}.
 \label{eq:appCmis}
\end{align}
For the RGB residual, let $\mathcal P_b$ average over $b\times b$ blocks, $b\in\{4,8,16\}$, and write $\bar Z_b=\mathcal P_b(FZ)/\mathcal P_b(F)$. At matched coverage,
\begin{equation}
 E_{\rm RGB}=\frac{1}{3\rho_0}\sum_b
 \left\langle\frac{\|\bar P_b-\bar A_b\bar I^\star_b\|_1}{3}
 \right\rangle_{\mathcal P_b(F)}.
 \label{eq:appCrgb}
\end{equation}

\ifdefined\backgroundLossColumnBreak\else\newpage\fi
\paragraph{Visible color-gain loss.}
Let $s_w=s(\mathbf1,x)$. Each string segment contributes to the gain $g_{\rm vis}$ in proportion to its visible share. The reference gain $\kappa$ uses the initial global palette $\{c_k^0\}$. Thus
\begin{align}
 g_{\rm vis}&=\sum_k V_k[s(c_k,x)-s_w]+V_o[s(\mathbf0,x)-s_w],\nonumber\\
 \kappa&=\max\{0,\max_k s(c_k^0,x)-s_w\}.
 \label{eq:appCgainfield}
\end{align}
For windows with $Q_{j\nu}=G_\nu*(D_j\kappa)>0$, the normalized local gain is $r_{j\nu}=G_\nu*(D_jg_{\rm vis})/Q_{j\nu}$. The loss penalizes a shortfall in this window aggregate relative to $\rho_B=\rho_0\min(1,\sqrt{B/1024})$. With $\softplus(z)=\log(1+e^z)$ and $T=0.1$,
\begin{equation}
 \mathcal L_{\rm gain}=
 \mathbb E_Q\!\left[
 \left(\frac{\softplus((1-r_{j\nu}/\rho_B)/T)}
              {\softplus(1/T)}\right)^{\!2}\right].
 \label{eq:appCgain}
\end{equation}
\ifdefined\backgroundLossColumnBreak\backgroundLossColumnBreak\fi
\paragraph{Background-suppression loss.}
We weight visible coverage by $1-F$ to penalize excess material in low-demand regions,
\begin{equation}
 \mathcal L_{\rm bg}=
 \frac{\sum_x(1-F(x))A(x)}{\max\{1,\sum_x(1-F(x))\}} .
 \label{eq:appCbg}
\end{equation}

\paragraph{Soft gate and string budget.}
For a legal pin pair $e$ in one color layer, let $d_e$ be its shortest pin-index distance to an active generatrix, accounting for periodic indexing and exchange of the two endpoints. Its Gaussian rendering weight is $\exp[-d_e^2/(2\sigma^2)]$. The hard selection radius is $w=\sigma=0.75$ pin-index units. Smoothing the radius test at temperature $\tau$ gives
\begin{equation}
 q_e=\frac{1}{1+\exp((d_e-w)/\tau)},\qquad
 \widetilde K=\sum_e q_e .
 \label{eq:implementedcount}
\end{equation}
We lower $\tau$ during training to sharpen the selection. The sum counts each filling pair once per layer and excludes the fixed outline. The budget loss is $\mathcal L_{\rm cnt}=((\widetilde K-B)/B)^2$.

\paragraph{Optimization.}
Adam refines the generatrices and colors. The same $q_e$ enters the binarization penalty averaged over candidate pairs in Sec.~4. During discrete re-search, we retain a locally refined proposal if it reduces $\mathcal L_{\rm total}$.

%% file: sec/D_loss_ablations.tex
\begingroup
\setcounter{figure}{0}
\renewcommand{\thefigure}{D.\arabic{figure}}
\renewcommand{\theHfigure}{D.\arabic{figure}}

\twocolumn[{%
  \begin{minipage}{\textwidth}
    \section{Loss-Function Ablations}
    \label{app:loss-ablations}
    \smallskip
    \input{figures/fig_loss_ablation_qualitative}
    \vspace{4pt}
  \end{minipage}%
}]

\paragraph{Color gain.}
Without the visible color-gain loss, white gaps open below the forewing tips of the first butterfly, where the target has orange markings (Fig.~\ref{fig:loss-ablation-qualitative}). Most of the blue tent canopy's center is also left white.

\paragraph{Mis-color.}
Without the mis-color loss, black segments cross the blue right half of the tent canopy. Dark segments also cross the pale wing bands of the butterfly in row~3.

\paragraph{Appearance.}
Without the appearance loss, widely spaced, nearly horizontal blue segments replace the crossing diagonals on the right side of the tent.

\ifdefined\lossAblationProofBreak\lossAblationProofBreak\fi

\paragraph{Count.}
Removing the count loss reduces the number of filling segments from 526 to 310 on the first butterfly and from 494 to 196 on the tent, leaving its cyan base sparsely covered.

\paragraph{Soft regularization.}
Without soft regularization, the blue segments on the right half of the tent form a slightly more pronounced fan.
\endgroup

%% file: figures/fig_loss_ablation_qualitative.tex
\captionsetup{type=figure}
\centering
\includegraphics[width=\linewidth]{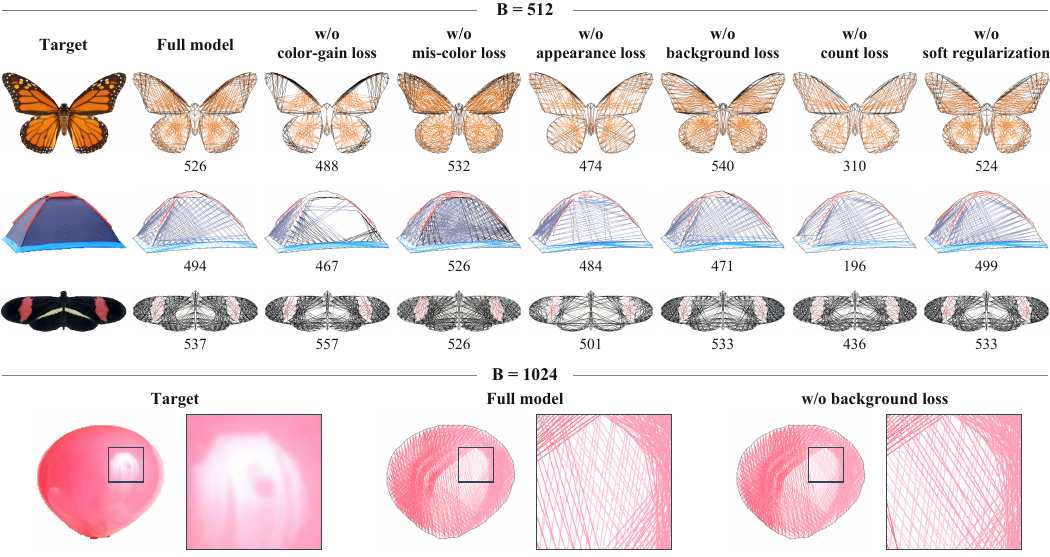}
\caption{\textbf{Loss-function ablations.}
$B$ is the target budget for filling string segments. The counts in rows 1--3 ($B=512$) exclude the fixed outline. The \emph{w/o soft regularization} setting removes both the gate-binarization and color-anchoring penalties.
At $B=1024$ (row~4), removing the background loss makes the pink string segments denser across the balloon's upper-right white highlight (insets).}
\label{fig:loss-ablation-qualitative}